\documentclass[proc]{edpsmath}
\usepackage{graphicx}
\usepackage[T1]{fontenc}
\usepackage[utf8]{inputenc}
\usepackage{verbatim}
\usepackage{subfig}
\usepackage{url}
\usepackage{stmaryrd}
\SetSymbolFont{stmry}{bold}{U}{stmry}{m}{n}
\usepackage{multirow}
\usepackage{caption} 

\newcommand{\nn}{\textbf{n}}
\newcommand{\FF}{\textbf{F}}
\newcommand\floor[1]{\lfloor#1\rfloor}

\graphicspath{{Images/}}

\usepackage{tikz}
\usetikzlibrary{shapes.arrows}

\newcommand{\varu}{\texttt{u}}
\newcommand{\Cell}{\texttt{Cell}}
\newcommand{\sizeof}{\texttt{sizeof}}
\newcommand{\RHS}{\texttt{RHS}}
\newcommand{\RHSLoc}{\texttt{RHS}}
\newcommand{\uLoc}{\texttt{uLoc}}
\newcommand{\NxLoc}{\texttt{NxLoc}}
\newcommand{\NyLoc}{\texttt{NyLoc}}
\newcommand{\NPart}{\texttt{NPart}}
\newcommand{\NPartX}{\texttt{NPartX}}
\newcommand{\NPartY}{\texttt{NPartY}}
\newcommand{\dFO}{\texttt{dataForOutput}}
\newcommand{\nDoF}{\texttt{nDoF}}
\newcommand{\nVar}{\texttt{nVar}}
\newcommand{\ovlpE}{\texttt{ovlpE}}
\newcommand{\ovlpN}{\texttt{ovlpN}}
\newcommand{\ovlpW}{\texttt{ovlpW}}
\newcommand{\ovlpS}{\texttt{ovlpS}}
\newcommand{\Read}{\textbf{R}}
\newcommand{\Write}{\textbf{W}}
\newcommand{\ReadWrite}{\textbf{RW}}
\newcommand{\F}{\boldsymbol{{F}}}
\newcommand{\FStar}{\textbf{F}^*}
\newcommand{\Nx}{\texttt{Nx}}
\newcommand{\Ny}{\texttt{Ny}}
\newcommand{\resp}{\textit{resp.~}}
\newcommand{\demi}{\frac{1}{2}}
\newcommand{\dx}{\Delta x}
\newcommand{\dy}{\Delta y}

\newcommand{\barU}{\bar{u}}
\newcommand{\barV}{\bar{v}}
\renewcommand{\u}{\boldsymbol{{u}}}
\newcommand{\n}{\boldsymbol{{n}}}
\newcommand{\un}{\u.\n}
\newcommand{\W}{\boldsymbol{W}}

\newcommand{\bitem}{\begin{itemize}}
\newcommand{\eitem}{\end{itemize}}
\newcommand{\benum}{\begin{enumerate}}
\newcommand{\eenum}{\end{enumerate}}

\newcommand{\tw}{\textwidth}
\newcommand{\beq}[1]{\begin{equation}\label{#1}}
\newcommand{\eeq}{\end{equation}}
\newcommand{\dd}[2]{\frac{\partial #1}{\partial #2}}
\newcommand{\Dt}{\mathcal{D}_t}
\newcommand{\KEvap}{\mathrm{K}}
\newcommand{\xv}{\vec{\boldsymbol{x}}}
\renewcommand{\u}{\boldsymbol{\vec{u}}}
\newcommand{\M}{\boldsymbol{M}}

\begin{document}

\title{A task-driven implementation of a simple numerical solver for hyperbolic conservation laws}
%

\author{Mohamed Essadki}
  \address{Laboratoire EM2C, CNRS, CentraleSupélec, Universit\'e Paris Saclay, Grande Voie des Vignes, 92295 Châtenay-Malabry - France}
  \secondaddress{IFP Énergies nouvelles, 1-4 avenue de Bois-Préau, 92852 Rueil-Malmaison Cedex - France}
\author{Jonathan Jung}
  \address{LMAP UMR 5142, UPPA, CNRS}
  \secondaddress{Cagire Team, INRIA Bordeaux Sud-Ouest, Pau - France}
\author{Adam Larat}
  \sameaddress{1}
  \secondaddress{Fédération de Mathématiques de l'École Centrale Paris, CNRS FR 3487, Grande Voie des Vignes, 92295 Châtenay-Malabry}
\author{Milan Pelletier}
  \sameaddress{1}
\author{Vincent Perrier}
  \sameaddress{4,\:3}

\begin{abstract} 
  This article describes the implementation of an all-in-one numerical procedure within the runtime StarPU. 
  In order to limit the complexity of the method, for the sake of clarity of the presentation of the non-classical 
  task-driven programming environnement, we have limited the numerics to first order in space and time. 
  Results show that the task distribution is efficient if the tasks are numerous and individually large enough 
  so that the task heap can be saturated by tasks which computational time covers the task management overhead. 
  Next, we also see that even though they are mostly faster on graphic cards,
  not all the tasks are suitable for GPUs, which brings forward the importance of the task scheduler. 
  Finally, we look at a more realistic system of conservation laws with an expensive source term, what allows us 
  to conclude and open on future works involving higher local arithmetic intensity, by increasing the order of 
  the numerical method or by enriching the model (increased number of parameters and therefore equations). 
\end{abstract}
\begin{resume}
  Dans cet article, il est question de l'implémentation d'une méthode numérique clé-en-main au sein du runtime StarPU.
  Afin de limiter la complexité de la méthode et ce dans le soucis de clarifier la présentation de notre méthode dans 
  le cadre de la programmation par tâche, nous avons limité l'ordre de la méthode numérique à un en temps et en espace. 
  Les résultats montrent que la distribution de tâches est efficace si les tâches sont suffisamment nombreuses et 
  de taille suffisante pour couvrir le temps supplémentaire de gestion des tâches. 
  Ensuite, nous observons que même si certaines tâches sont nettement plus rapides sur carte graphique, 
  elles ne sont pas toutes éligibles à un portage sur GPU, ce qui met en avant 
  l'importance d'un répartiteur de tâche intelligent. 
  Enfin, nous regardons un système de lois de conservation plus tournée vers l'application, 
  incluant notamment un terme source co\^uteux en terme de temps de calcul. 
  Ceci nous permet de conclure et d'ouvrir sur un travail futur, dans lequel l'intensité arithmétique locale sera 
  augmentée par le biais d'une méthode numérique ou d'un modèle d'ordres plus élevés. 
\end{resume}

\maketitle

\section{Introduction}
\label{sec:Intro}

Computations of turbulent flows has been undergoing two breakthrough over the last three decades,
concerning computational architectures and numerical methods. On the one hand,  computing clusters are
reaching a size of many hundreds of thousands of cores, which nowadays allows to overcome 
RANS modeling by considering unstationary LES simulations \cite{RefLES1,RefLES2}. On the other hand, this new framework requires
high order numerical methods, which have also been much improved recently \cite{RefLES1,ZhangXiaShu12,DumbserBalsaraToro08}. 
Nevertheless, these newly manufactured massively multicore computers rely on heterogeneous architectures.
In order to improve the efficiency of our methods on these architectures, their implementation and
algorithmic must be redesigned. The aim of this project is to optimize the 
implementation of all-in-one numerical methods and models on emerging heterogeneous 
architectures by using runtimes schedulers. In this publication, we focus on StarPU \cite{augonnet}.

Runtime schedulers allow to handle heterogeneous architectures in a transparent way. 
For this, the algorithm must be recast in a graph of tasks. 
Computing kernels must then be written in different languages, in order to be executed
on the different possible devices. Then, the runtime scheduler will be in 
charge of dynamically balancing the tasks on the different available computing 
resources (cores of the host, accelerators, etc.).

Finite differences methods have already been successfully ported on hybrid architectures \cite{RefFDHybrid}.
However, we aim at developping numerics on unstructured meshes, for example of the discontinuous Galerkin type, 
and our methods have a very different memory access pattern, which is one of the bottlenecks of the task-driven 
programming. In order to minimize the complexity of our presentation, gathering both the numerical procedure 
and its implementation within the non-classical StarPU framework, we have decided to restrict its scope to first order in space and time.

\section{A first order finite volume solver for two dimensional conservation laws}
\label{sec:FV0}
\subsection{Hyperbolic conservation laws}
\label{ssec:HCL}

Let $\Omega$ be a subset of $\mathbb{R}^2$.  
We consider a two dimensional system of conservation laws with source terms
\begin{equation}
\label{eq:conservation_law}
\partial_t W + \nabla \cdot \FF(W)= S(W),
\end{equation}
where $t$ is the time, $\nabla=(\partial_x,\partial_y)^T$ with $(x,y)$ the spatial position,  
$W\in\mathbb{R}^m$ ($m\in\mathbb{N}$) is the vector of conservative variables, 
$\FF=(F_x,F_y)^T:\Omega \rightarrow \mathbb{R}^m$ is the flux function and $S(W)$ 
represents the source terms. Later, the number of conserved variables is noted $\nVar = m$. 
The unknowns $W$ depend on the spatial position $(x,y)$ and on the time $t$. 
We assume that system~\eqref{eq:conservation_law} is hyperbolic, meaning that the directional Jacobian
$\dd{\FF}{W}\cdot\nn$ is diagonalizable in every direction $\nn\in\mathcal{S}^1$, 
and we note $\lambda_1(W)\leq \cdots\leq\lambda_m(W)$ its eigenvalues, the direction $\nn$ being implicit.
We aim at computing a solution of \eqref{eq:conservation_law} with initial 
condition
\begin{equation}
\label{eq:initial_condition}
\forall (x,y)\in\Omega,\quad W(x,y,0)=W^0(x,y).
\end{equation}

\subsection{Finite volume discretization}
\label{ssec:FVDiscr}

In this section, we present a simple numerical scheme which solves system \eqref{eq:conservation_law} in space and time: $(x,y,t)\in\Omega\times\mathbb{R}^+.$
If $\Nx$ and $\Ny$ are two given positive integers, we define the sequences $\left(x_{i+\frac{1}{2}}\right)_{0\leq i\leq \Nx}$ and 
$\left(y_{j+\frac{1}{2}}\right)_{0\leq j\leq \Ny}$ by $x_{i+\frac{1}{2}}=a+i\Delta x,\: i=0,\cdots, \Nx$ and $y_{j+\frac{1}{2}}=c+j\Delta y,\: j=0,\cdots, \Ny$, 
so that $\Omega$ is discretized into  $\Nx\times \Ny$ cells $C_{i,j}=\left]x_{i-\frac{1}{2}},x_{i+\frac{1}{2}}\right[\times\left]y_{j-\frac{1}{2}},y_{j+\frac{1}{2}}\right[$. 
We also  consider a sequence of times $(t_n)_{n\in\mathbb{N}}$ such that $t_0=0$, which defines the time steps $(\Delta t)_n=t_{n+1}-t_n>0$.

\subsubsection{Hyperbolic part}\label{sssec:FV}

Next, for any cell $C_{i,j}$ of the mesh, at any time $t_n$, we look for an approximation of the average value of the solution $W(x,y,t)$ on the cell:
\beq{eq:Average}
W_{i,j}^n=\frac{1}{\vert C_{i,j}\vert}\int_{C_{i,j}} W(x,y,t_n) dx dy.
\eeq
By integrating system \eqref{eq:conservation_law} over a cell $C_{i,j}$, one obtains:
\begin{align*}
\frac{d}{dt} \int_{C_{i,j}} W(x,y,t) dxdy +\sum_{\Gamma\subset\partial C_{i,j}} \int_{\Gamma} \FF(W)\cdot \nn \: dS=0,
\end{align*}
where $\nn=(n_x,n_y)$ is the outward unit normal to the boundary $\partial C_{i,j}$.
Then, the first order explicit finite volume scheme writes \cite{leveque02}:
\begin{align}
W_{i,j}^{n+1,*}-W_{i,j}^n=-\frac{(\Delta t)_n}{\Delta x} \left(\tilde{F}\left(W_{i,j}^n,W_{i+1,j}^n,(1,0)\right)-\tilde{F}\left(W_{i-1,j}^n,W_{i,j}^n,(1,0)\right)\right)\nonumber\\
- \frac{(\Delta t)_n}{\Delta y} \left(\tilde{F}\left(W_{i,j}^n,W_{i,j+1}^n,(0,1)\right)-\tilde{F}\left(W_{i,j-1}^n,W_{i,j}^n,(0,1)\right)\right),
\label{eq:VF_scheme}
\end{align}
where $\tilde{F}(W_{k,l},W_{m,n},\nn)$ is a chosen numerical flux. In all our computations, 
the first order Lax-Friedrichs flux will be used: 
$$\tilde{F}(W_{L},W_{R},\nn)
 = \dfrac{ \FF(W_{L}) \cdot \nn + \FF(W_{R}) \cdot \nn }{2} - 
\dfrac{\sigma}{2} \, \left( W_R - W_L\right)
$$
with 
$$\sigma = 
\underset{p\in\llbracket 1,m\rrbracket}{\max} \left( \vert\lambda_p(W_{L})\vert \, , \,  
\vert\lambda_p(W_{R})\vert \right).
$$
The numerical scheme \eqref{eq:VF_scheme} is stable provided the following 
CFL condition is ensured 
\begin{equation}
\label{eq:CFL_cond}
(\Delta t)_n \times \underset{(i,j)\in \llbracket 1,\Nx\rrbracket\times\llbracket 1,\Ny\rrbracket}{\max} \quad\left(\underset{p\in\llbracket 1,m\rrbracket}{\max}\vert\lambda_p(W_{i,j}^n)\vert\right) \leq \min(\Delta x,\Delta y).
\end{equation}
In practice, a constant time step $(\Delta t)_n = \Delta t$ is set at start, 
and therefore we just need to check at the beginning of each iteration that 
$\Delta t$ satisfies inequality \eqref{eq:CFL_cond}.

\subsubsection{Source term}\label{sssec:Source}

Since we consider only a first order numerical scheme in space and time, the source term 
treatment can be simply added by use of a first order operator splitting technique \cite{HV03}.

Therefore, the additional source terms can possibly be taken into account by local 
cell-wise integration of the following ODE, which is simply discretized by mean of 
a forward Euler time scheme: 
\beq{eq:SourceTerm}
  \partial_t W = S(W) 
  \qquad \longmapsto \qquad
  W_{i,j}^{n+1} = W_{i,j}^{n+1,*} + \Delta t S\left(W_{i,j}^{n+1,*}\right),
\eeq
where $W_{i,j}^{n+1,*}$ is the partial update given by the Finite Volume integration 
\eqref{eq:VF_scheme}. Let us note that though the update \eqref{eq:SourceTerm} is 
simple and cheap, the computation of $S\left(W_{i,j}^{n+1,*}\right)$ may be rather 
complex and computationnally expensive. 

%

\section{StarPU: a task scheduler}

Task scheduling offers a new approach for the implementation of numerical codes, which is 
particularly suitable when looking for an execution on heterogeneous architectures.

The structure of the codes is divided into two main layers that we will call here 
\textit{bones} and \textit{flesh}. At first level, the code is parsed to build a 
tasks dependency graph that can be viewed as its \textit{skeleton}. 
Next, one enters within this graph of tasks and starts executing the actions associated 
to each of these tasks. Then, the code acts on the memory layout thanks to what can be 
viewed as its \textit{muscles}. During the execution part, one may have different choices 
in the tasks execution on the available computational power and this is where the 
\texttt{scheduler} plays its role: it aims at distributing the actions so as to minimize 
the global computational time.

Now, we detail the glossary attributed to such a new implementation framework within 
the context of the StarPU runtime.

\label{sec:StarPU}
\subsection{StarPU tasks}
\label{sec:StarPUGlossary}
In StarPU, a task is made of the following: a \texttt{codelet}, associated \texttt{kernels} and \texttt{data handles}.
\bitem
  \item The \texttt{codelet} is the task descriptor. 
    It contains numerous information about the task, including the number of data handles, 
    the list of available kernels implemented to execute this task 
    (for CPU, GPU or possibly something else\dots) 
    and the memory access mode for each data handle: 
    "Read" (R), "Write" (W) or "Read and Write" (RW).
  \item The \texttt{kernels} are the functions that will be executed on a dedicated architecture: 
    CPU, GPU, etc. A task may have the choice between different kernels implementation and it 
    is the role of the StarPU scheduler to distribute the tasks on the available heterogeneous 
    architectures following a certain criterion 
    (in general minimizing the global execution time). 
  \item The \texttt{data handles} are the memory managers. Each data handle can be viewed 
    as the encapsulation of a memory allocation, which allows to keep trace of the action 
    (read or/and write) of the task kernel on the memory layout. In particular, 
    this allows to build the task dependency graph. 
\eitem

\subsection{Construction of the tasks dependency graph}

The domain of size $\Nx\times\Ny$ is partitioned 
into $\NPartX\times\NPartY$ balanced parts of size $\NxLoc\times\NyLoc$, 
see Figure \ref{fig:partition}. 
One task will deal with one partition at a time. 
In order to treat the interaction between the domains in an asynchronous manner, 
each domain is supplemented with copies of its border data, see Figure \ref{fig:overlap}. 

\begin{figure}
\centering
\begin{minipage}[b]{0.73\tw}
\hspace{-.6cm}
\resizebox{1.02\tw}{!}{%
\begin{tikzpicture}
\draw[step=0.2,black,thin,fill=blue!40] (0.,0.) grid (6,6) rectangle(0,0);

\coordinate (a) at (6.5,3);
\coordinate (b) at (8.2,3);
\draw[->, >=latex, blue, line width=5pt]   (a) -- (b) ;

\draw[step=0.2,black,thin,fill=yellow!40,xshift=8.4cm,yshift=-0.2cm] (0.,0.) grid (3,3) rectangle(0,0);

\draw[step=0.2,black,thin,fill=red!40,xshift=8.4cm,yshift=3.2cm] (0.,0.) grid (3,3) rectangle(0,0);

\draw[step=0.2,black,thin,fill=pink!40,xshift=11.8cm,yshift=-0.2cm] (0.,0.) grid (3,3) rectangle(0,0);

\draw[step=0.2,black,thin,fill=green!40,xshift=11.8cm,yshift=3.2cm] (0.,0.) grid (3,3) rectangle(0,0);

\end{tikzpicture}
}%
\captionsetup{font=footnotesize}
\captionof{figure}{\label{fig:partition} Partitioning of an initial mesh of 
$\left\{\Nx=30\times \Ny=30 \right\}$ cells
into $\left\{\NPartX=2 \times \NPartY=2 \right\}$ 225 cells domains. }
\end{minipage}%
\begin{minipage}[b]{0.4\tw}
\resizebox{0.9\tw}{!}{%
\begin{tikzpicture}
\draw[step=0.2,black,thin,fill=blue!40] (0.,0.) grid (3,3) rectangle(0,0);
\draw[step=0.2,black,thin,fill=red!40,yshift=3.2cm] (0.,0.) grid (3,0.2) rectangle(0,0);
\draw (1.5,3.4) node[above] {North};
\draw[step=0.2,black,thin,fill=red!40,yshift=-0.4cm] (0.,0.) grid (3,0.2) rectangle(0,0);
\draw (1.5,-0.4) node[below] {South};
\draw[step=0.2,black,thin,fill=red!40,xshift=3.2cm] (0.,0.) grid (0.2,3) rectangle(0,0);
\draw (3.4,1.5) node[right] {East};
\draw[step=0.2,black,thin,fill=red!40,xshift=-0.4cm] (0.,0.) grid (0.2,3) rectangle(0,0);
\draw (-0.4,1.5) node[left] {West};
\end{tikzpicture}
}
\captionsetup{font=footnotesize}
\captionof{figure}{\label{fig:overlap} Partition (in blue) with its overlap (in red) 
in the East, North, West and South direction. Data of each partition is composed of 
these five handles.}
\end{minipage}%
\end{figure}

\bitem
  \item Each memory allocation is associated with a \texttt{data handle}: 
    one for each subdomain, four for its borders copy, called \textit{overlaps} in the following, 
  \item The \texttt{codelet} of each task points toward a certain number of these data handles, 
    tagged as "R" or "W" or "RW", for  \textit{Read} and/or \textit{Write}, following the access mode exepected by the 
    associated \texttt{kernel},  
  \item Finally, the code is parsed and the tasks are submitted to StarPU. 
    When a task has to \emph{write} on a certain data handle memory, 
    a dependency edge is drawn between this task and the latest tasks having \emph{read} access on this data handle. 
    Similarly, for each of its \emph{read-accessed} data handles, 
    a task will depend on the latest tasks having \emph{write} access on them.
  \item Once the dependency graph is built, the tasks are distributed in dependency order by the scheduler on the 
    available computational ressources, following the available kernel implementations. Note that building the dependency 
    graph and lauching the task can be an intricated work, especially since the future of the graph may depend on 
    the computation itself: think of the number of time steps which may depend on the solution, for example. 
\eitem

\subsection{Task overhead}
\label{ssec:Overhead}
\begin{figure}
\begin{center}
		\subfloat[\small 2 dodeca-core Haswell Intel Xeon E5-2680.]{
			\begin{minipage}[c]{0.5\textwidth}
				\includegraphics[width=1.0\tw]{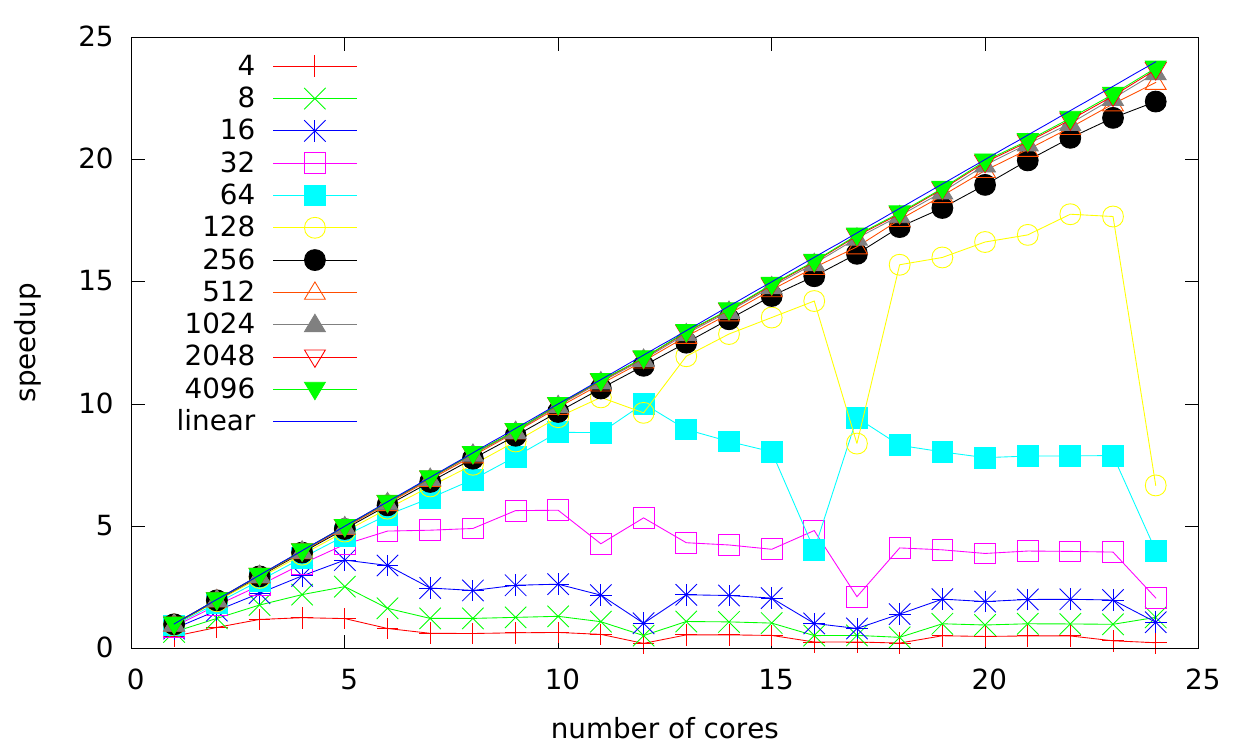}
			\end{minipage}
    }
		\subfloat[\small Xeon Phi KNL.]{
			\begin{minipage}[c]{0.5\textwidth}
				\includegraphics[width=1.0\tw]{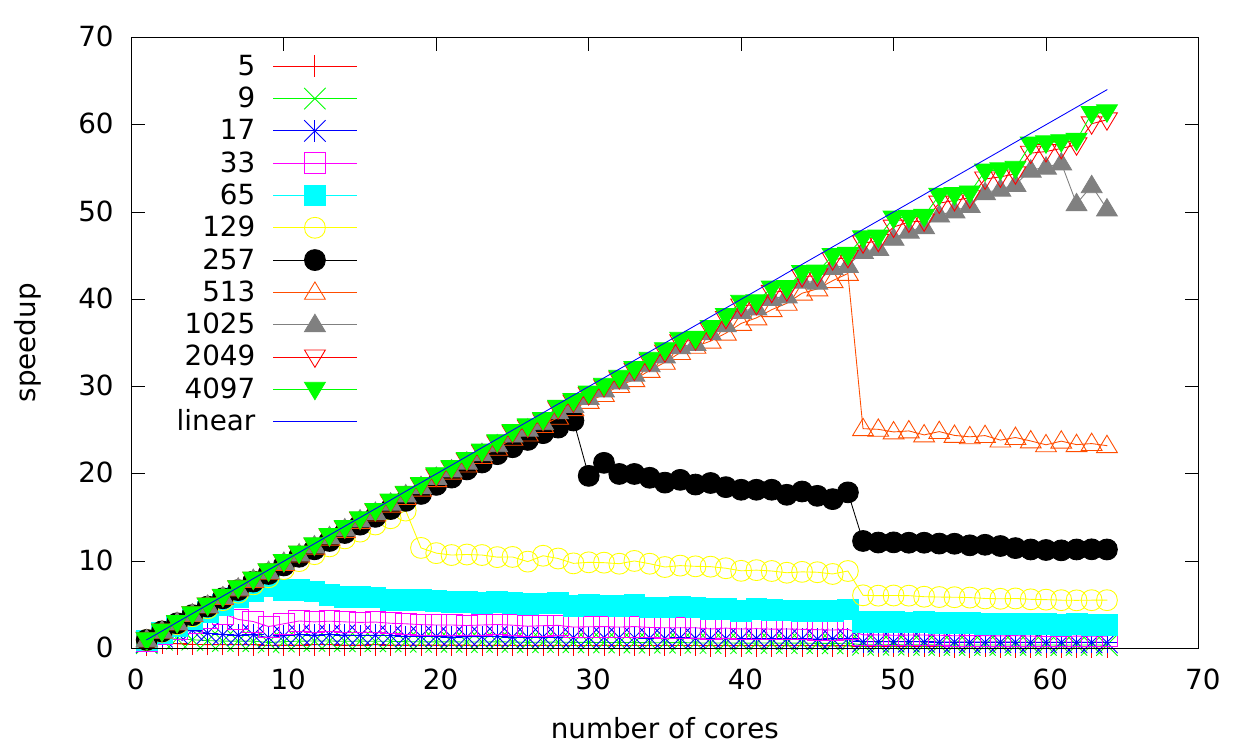}
			\end{minipage}}
		\caption{Task size overhead with the \texttt{eager} scheduler: scalability results obtained with duration of tasks varying between $4$ and $4096\:\mu s$ on two different machines.}
		\label{fig:TaskSizeOverheadEager}
	\end{center}
\end{figure}

StarPU tasks management is not free in terms of CPU time. In fact, each task execution presents 
a latency called "\textit{overhead}". 
We want to know when this additional time can be neglected or not. 
There is a StarPU benchmark that allows to measure the minimal duration of a task to ensure a good scalibility. We send $1000$ short tasks with the same (short) duration varying between $4$ and $4096\:\mu s$  and we study the scalability. In Figure \ref{fig:TaskSizeOverheadEager}, we plot the results obtained on a 2 dodeca-core Haswell Intel Xeon E5-2680 and on a Xeon Phi KNL with the scheduler \texttt{eager}. On few cores, we have a good scalibility result, even if the duration of the tasks is short ($<0.2 ms$). On many cores, we need longer tasks duration to get satisfying scaling ($\approx 1 ms$).

To sum up, if the duration of the tasks is smaller than the microsecond, their overhead cannot be neglected anymore. 

\subsection{Schedulers}
\label{sec:schedulers}
The purpose of the scheduler is to launch the tasks when they become ready to be executed. 
In StarPU, many different scheduling policies are available. 
In this paper, we consider only the two following:
\bitem
  \item the \texttt{eager} scheduler: it is the scheduler by default. It uses a central task queue. As soon as a worker has finished a task, the next task in the queue is submitted to this worker.
  \item the \texttt{dmda} scheduler: this scheduler takes into account the performance models and the data transfer time
   for the tasks (see Section \ref{subsec:PerfModel}). 
   It schedules the tasks so as to minimize their completion time by carefully choosing the optimal executing architecture.  
\eitem

\section{Practical Implementation}
\label{sec:PracticalImplementation}

After a general presentation of the model and the numerical method in section \ref{sec:FV0} 
and of the runtime environment with a particular focus on StarPU in section \ref{sec:StarPU}, 
we now detail the way we have implemented this numerical resolution of a system of hyperbolic 
conservation laws within the task-driven framework StarPU. 

\subsection{Memory allocation}

Since the tasks dependency graph is mainly based on the memory dependency between successive 
tasks, memory allocation is a crucial development part of our application. 
This starts with the definition of the structure \Cell, which contains the $\nVar=m$~
conserved variables at a certain point in space and time. 

\subsubsection{Principal variables}

For a first order finite volume resolution of a system of conservation laws, only two 
main variables are needed:
\bitem
  \item $\varu$, the \textbf{computed solution}. In each mesh cell, it contains 
    $\nDoF\times\nVar$ floats. However, since at first order $\nDoF = 1$, \varu~contains 
    only one \Cell~structure of size \nVar~corresponding to one local approximation \eqref{eq:Average} of the solution per mesh cell.
  \item \RHS, the vector of \textbf{residuals}. It has exactly the same memory 
    characteristics as \varu, since at each time step, the update is done by:
    \beq{eq:update}\varu \; +\!\!= \; \RHS.\eeq
\eitem

These two vectors of variables are allocated per subdomain. 
Typically, for each of the $\NPartX\times\NPartY$ subdomains, a vector \uLoc~of 
$(\NxLoc\times\NyLoc)$ \Cell s~is created and encapsulated in a StarPU data handler which allows to 
follow the memory dependency.  

\subsubsection{Overlap additional memory buffers}

In order to minimize the communications and the dependencies between subdomains, 
each subdomain comes with four additional one-dimensional buffers corresponding to 
the possible four overlap-data needed at the subdomain boundaries (East, North, West and South). 

Two vectors of size $\NxLoc\times\sizeof(\Cell)$ (\ovlpS~and \ovlpN) and 
two vectors of size $\NyLoc\times\sizeof(\Cell)$ (\ovlpE~and \ovlpW) are always allocated, 
whether the overlap needs to be used or not. The reason for that is that the number of 
data handlers passed to a StarPU kernel needs to be constant. Therefore, when copying the overlap-data 
for example, all the overlap data handlers are passed anyway but nothing is done if they are not 
needed, like in the case of a periodic subdomain in one direction. 

Of course, here lies a small communication optimization, when sending a task on another device, since 
some useless data is transfered. However, we think that the overlap tasks should be of negligeable 
size compared to the task acting on the entire subdomains and should be mainly executed on the host node.

\subsection{Description of the tasks}
\label{sec:Tasks_description}
In paragraph \ref{sssec:FV}, we have seen that the first order finite volume method 
essentially consists in 
1) checking the computational time step $\Delta t$, 
2) computing all the numerical fluxes at all the edges of the mesh, 
3) gathering them in the \RHS~vector 
and finally 
4) updating the numerical solution \varu.

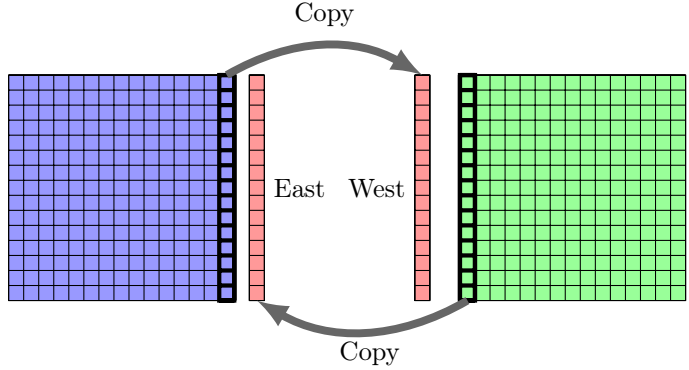
\begin{figure}
\begin{tikzpicture}


\draw[step=0.2,black,thin,fill=blue!40] (0.,0.) grid (3,3) rectangle(0,0);

\draw[step=0.2,black,ultra thick] (2.8,0.) grid (3,3) rectangle(2.8,0.);

\draw[step=0.2,black,thin,fill=red!40,xshift=3.2cm] (0.,0.) grid (0.2,3) rectangle(0,0);
\draw (3.4,1.5) node[right] {East};


\draw[step=0.2,black,thin,fill=green!40,xshift=6cm] (0.,0.) grid (3,3) rectangle(0,0);

\draw[step=0.2,black,ultra thick,xshift=6cm] (0.,0.) grid (0.2,3) rectangle(0.,0.);

\draw[step=0.2,black,thin,fill=red!40,xshift=5.4cm] (0.,0.) grid (0.2,3) rectangle(0,0);
\draw (5.4,1.5) node[left] {West};

\draw[->, >=latex,black!60, line width=3pt]   (2.9,3) to[bend left] (5.5,3) ;
\draw (4.2,3.8) node  {Copy};

\draw[->, >=latex, black!60, line width=3pt]   (6.1,0) to[bend left] (3.3,0) ;
\draw (4.8,-0.7) node  {Copy};

\end{tikzpicture}

\caption{\label{fig:overlapcommunications}
Copy to overlaps tasks, example 
with two domains. The global computational domain is vertically divided into 
two parts, blue (left) and green (right). 
The green one is supplemented with a west overlap, whereas the blue one is 
supplemented with an east overlap. One residual computation includes two 
communications tasks: copying the right column of the blue domain into the west overlap of the 
green domain, and copying the left column of the green domain into the east overlap of the blue domain.}
\end{figure}

\begin{figure}[ht!!]
  \centering
  \includegraphics[width=1.0\tw]{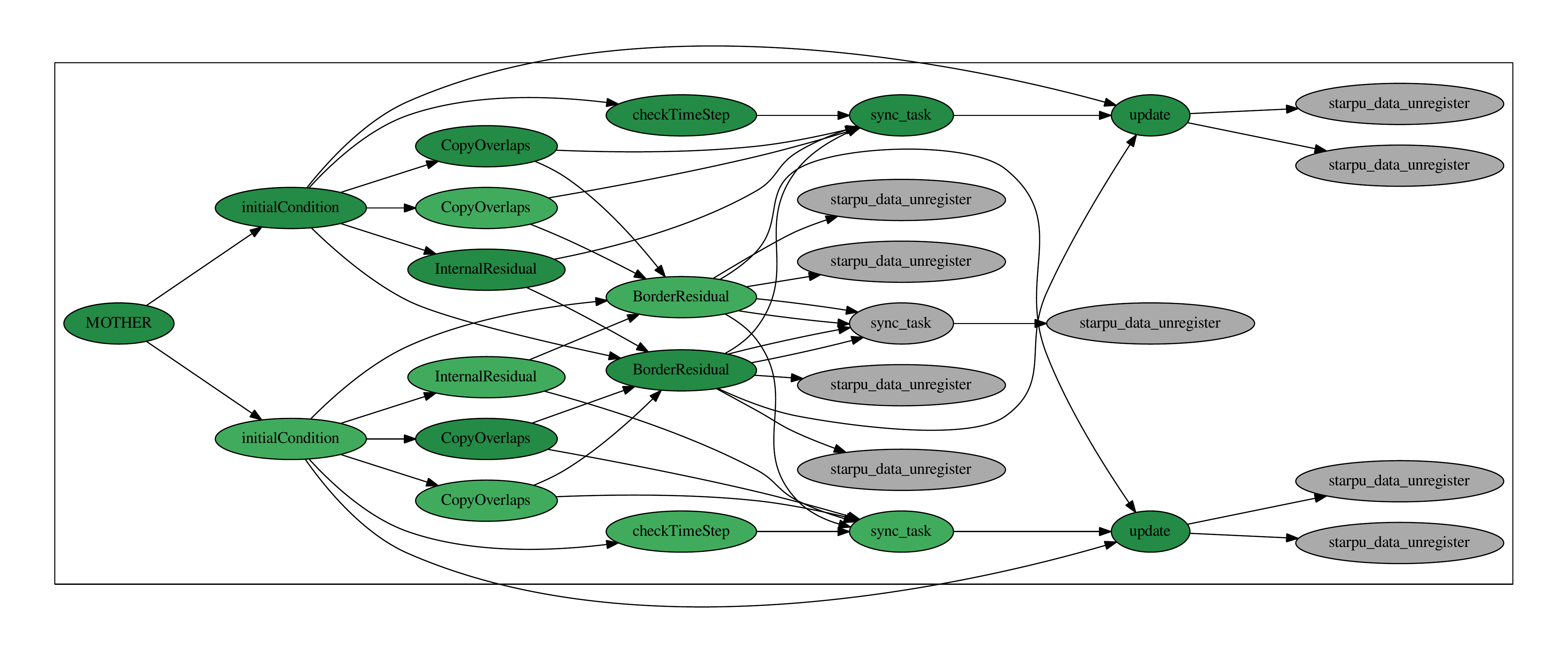}
  \caption{\label{fig:TaskDiagramNoOutput} Task diagram built by StarPU for one Forward Euler step
    of our first order finite volume task-driven implementation on two subdomains, hence the horizontal symmetry.}
\end{figure}

Based on this general decomposition, here is the list of tasks implemented in our application.
Between brackets is specified the memory data handlers accessed by the task, with their respective  
access rights given between parentheses:
\bitem
  \item \texttt{initialCondition}\big[\uLoc(\Write)\big]: fills each subdomain solution with 
    the initial condition.  
  \item \texttt{checkTimeStep}\big[\uLoc(\Read)\big]: computes the largest characteristic speed within
    the subdomain. In order to avoid gathering these time constraints globally, we only check that the fixed time step $\Delta t$ initially given by the user respects locally the stability constraint.
  \item \texttt{copyOverlaps}\big[\ovlpE(\Write),\ovlpN(\Write),\ovlpW(\Write),\ovlpS(\Write),\uLoc(\Read)\big]: 
    each subdomain fills the corresponding neighbors overlap data vectors, if needed. This is not
    the case if the subdomain is periodic in one or both directions, or if one of the boundary states 
    is prescribed (Dirichlet boundary condition). The northest (\resp southest) 
    line of the subdomain is copied in the \ovlpS~(\resp \ovlpN) data handle of the 
    northern (\resp southern) neighbor. The extreme east (\resp extreme west) column is copied in the 
    \ovlpW~(\resp \ovlpE) data handle of the eastern (\resp western) neighbor. 
    Figure \ref{fig:overlapcommunications} depicts the \texttt{copyOverlaps}
    task for two domains. 
  \item \texttt{internalResiduals}\big[\uLoc(\Read),\RHSLoc(\Write)\big]: computes the numerical flux
    at all the edges of the subdomain, except those of the boundaries where an overlap data vector 
    has to be used: 
    \beq{eq:RHSFilling}
      \RHS_{i,j} = \frac{\Delta t}{\dx}\left(\FStar_{i+\demi,j}-\FStar_{i-\demi,j}\right)
                 + \frac{\Delta t}{\dy}\left(\FStar_{i,j+\demi}-\FStar_{i,j-\demi}\right).
    \eeq
  \item \texttt{borderResiduals}\big[\ovlpE(\Read),\ovlpN(\Read),\ovlpW(\Read),\ovlpS(\Read),\uLoc(\Read),\RHSLoc(\ReadWrite)\big]: computes the remaining numerical fluxes, meaning those between 
    the overlap vector states and the border cells of the subdomain. Therefore, the overlap data 
    vectors passed in argument here are those belonging to the subdomain. 
  \item \texttt{update}\big[\uLoc(\ReadWrite),\RHSLoc(\Read)\big]: update the numerical solution 
    subdomain-wise, thanks to the update relation \eqref{eq:update}.
\eitem
The corresponding task diagram for one time step and two sub-domains is given in Figure 
\ref{fig:TaskDiagramNoOutput}. This graph is an output we can get from StarPU to verify 
the correct sequence of the tasks. 

\subsection{Specific asynchronous treatment of the outputs}

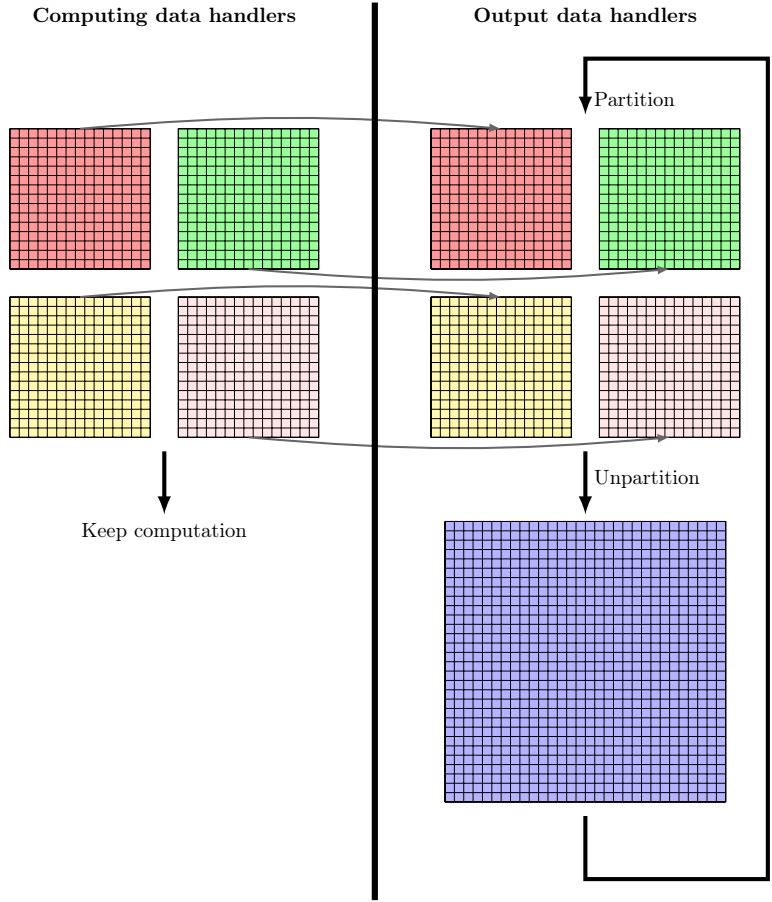
\begin{figure}
\resizebox{0.6\tw}{!}{%
\begin{tikzpicture}[scale=0.8]

  \def\xsep{9cm}
  \def\lrect{3cm}
  \def\ssep{0.3cm}


  \draw (\lrect+\ssep,2*\lrect+10*\ssep) node  
        {\textbf{Computing data handlers}};

  \draw[step=0.2,black,thin,fill=yellow!40] (0.,0.) grid (\lrect,\lrect) rectangle(0,0);
  \draw[step=0.2,black,thin,fill=red!40,yshift=\lrect+2*\ssep] (0.,0.) grid (\lrect,\lrect) rectangle(0,0);
  \draw[step=0.2,black,thin,fill=pink!40,xshift=\lrect+2*\ssep] (0.,0.) grid (\lrect,\lrect) rectangle(0,0);
  \draw[step=0.2,black,thin,fill=green!40,xshift=\lrect+2*\ssep,yshift=\lrect+2*\ssep] (0.,0.) grid (\lrect,\lrect) rectangle(0,0);

  \draw[->, >=latex, black, line width=2pt]   (\ssep+\lrect,-\ssep) to (\ssep+\lrect,-5.5*\ssep) node[below] {Keep computation};


  \draw (\lrect+\ssep+\xsep,2*\lrect+10*\ssep) node  
        {\textbf{Output data handlers}};

  \draw[step=0.2,black,thin,fill=yellow!40,xshift=\xsep] (0.,0.) grid (\lrect,\lrect) rectangle(0,0);
  \draw[step=0.2,black,thin,fill=red!40,xshift=\xsep,yshift=\lrect+2*\ssep] (0.,0.) grid (\lrect,\lrect) rectangle(0,0);
  \draw[step=0.2,black,thin,fill=pink!40,xshift=\lrect+2*\ssep+\xsep] (0.,0.) grid (\lrect,\lrect) rectangle(0,0);
  \draw[step=0.2,black,thin,fill=green!40,xshift=\lrect+2*\ssep+\xsep,yshift=\lrect+2*\ssep] (0.,0.) grid (\lrect,\lrect) rectangle(0,0);

  \draw[->, >=latex,black!60, line width=1pt]   (0.5*\lrect,2*\lrect+2*\ssep) to[bend left=5] (0.5*\lrect+\xsep,2*\lrect+2*\ssep) ;
  
  \draw[->, >=latex,black!60, line width=1pt]   (0.5*\lrect,\lrect) to[bend left=5] (0.5*\lrect+\xsep,\lrect) ;

  \draw[->, >=latex,black!60, line width=1pt]   (0.5*\lrect+\lrect+2*\ssep,\lrect+2*\ssep) to[bend right=5] (0.5*\lrect+\lrect+2*\ssep+\xsep,\lrect+2*\ssep) ;

  \draw[->, >=latex,black!60, line width=1pt]   (0.5*\lrect+\lrect+2*\ssep,0) to[bend right=5] (0.5*\lrect+\lrect+2*\ssep+\xsep,0) ;

  \draw[->, >=latex, black, line width=2pt]   (\xsep+\ssep+\lrect,-\ssep) to (\xsep+\ssep+\lrect,-5.5*\ssep) node[above right,yshift=\ssep] {Unpartition};

  \draw[step=0.2,black,thin,fill=blue!30,xshift=\xsep+\ssep,yshift=-2*\lrect-6*\ssep] (0.,0.) grid (2*\lrect,2*\lrect) rectangle(0,0);


  \draw[->, >=latex, black, line width=2pt,yshift=-2*\lrect-6*\ssep]   (\xsep+\ssep+\lrect,-\ssep) -- (\xsep+\ssep+\lrect,-5.5*\ssep) -- (\xsep+4*\ssep+2*\lrect,-5.5*\ssep) -- (\xsep+4*\ssep+2*\lrect,5*\lrect + 3*\ssep) -- (\xsep+\ssep+\lrect,5*\lrect + 3*\ssep) -- (\xsep+\ssep+\lrect,5*\lrect -\ssep) node[above right] {Partition};

  \draw[black, line width=3pt]   (0.5*\xsep+\lrect+\ssep,2*\lrect+11*\ssep) to (0.5*\xsep+\lrect+\ssep,-2*\lrect-13*\ssep) ;

\end{tikzpicture}
}%
\caption{\label{fig:partitionIO} 
  Strategy for non blocking IO. The computing buffers are depicted on the left. 
  The data handlers for output \dFO~are depicted on the right. Each 
  part of the domain is writing on a dedicated part of \dFO~using the 
  task \texttt{gatherForOutput}. Once this copy 
  is done, the computing buffers can be used for further computations. 
  On the output handlers side, once all the copies have been completed, 
  the buffer can be switched to a global buffer thanks to a \texttt{switch} 
  task, which is followed by a 
  \texttt{starpu\_data\_invalidate\_submit} command on the local 
  descriptor of \dFO. Once the unpartitioning is done,  
  the output handler can be used as a 
  single data handler, and \emph{one} tasks is in charge of writing this 
  buffer in an output file. Once the ouput in the file is finished, another 
  \texttt{switch} task is in charge of re-partitioning the data, the 
  global descriptor is invalidated, and the \dFO~is able to receive data 
  from the computing buffers again. 
}

\end{figure}

The output task consists in writing the global mesh solution \varu~into a disk file
at once. In order to avoid global synchronization, 
a data handle \dFO~of size $\Nx\times\Ny\times\nVar$ is used to temporarily 
store the data of each subdomain. Since this buffer will be entirely written in an output file 
by a single task \texttt{outputTask}, it needs to be contiguous in memory and to be 
encapsulated in a single global descriptor. 
However, when the solution needs to be written on the disk, 
it is first copied in this \dFO~buffer and this should be done in an asynchronous parallel manner, 
by the \texttt{gatherForOutput} tasks. 
This is only possible if the \dFO~buffer is also described per subdomain, see Figure~\ref{fig:partitionIO}.
At the end, we need one global data handle and a set of $\NPartX\times\NPartY$ 
per-subdomain data handlers, both sharing the same memory allocation. 
Since the dependency graph is built on the memory dependancy between the tasks, 
it is very dangerous to use data handlers sharing the same memory slots. 


Fortunately, both descriptors can be declared in StarPU and a specific procedure allows to switch 
from one to another, so that they are never used concurrently. 
First, the global buffer is allocated and encapsulated in a global 
StarPU data handler. Next, it is also described as a vector of $\NPartX \times \NPartY$ data 
handlers, thanks to the command:
\begin{verbatim}
  starpu_matrix_data_register(starpu_data_handle *data_handle,int DEVICE,void *data_ptr,
                                int LD,int NxLoc, int NyLoc, size_t sizeof(Cell));
\end{verbatim}
Here, the integer \texttt{LD} allows to encapsulate the grid block per block, since the 
global vector is accessed through the formula:
\begin{verbatim}
  for(int j=0;j<NyLoc;j++) {for(int i=0;i<NxLoc;i++) global_data[i+j*LD];}
\end{verbatim} 
Then, if \texttt{data\_ptr} points to the bottom-left corner of the subdomain, the corresponding block
is encapsulated in the subdomain data handler \texttt{data\_handle}. 

\begin{figure}[ht!!]
  \centering
  \includegraphics[width=1.0\tw]{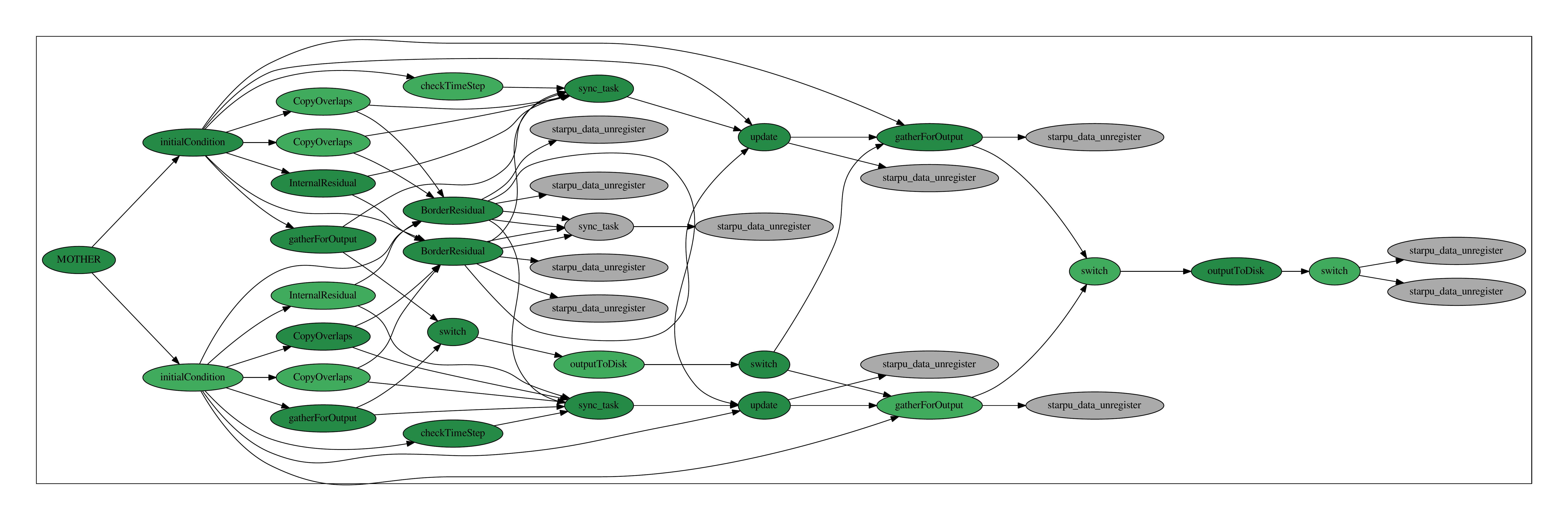}
  \caption{\label{fig:TaskDiagramWithOutput} Task diagram built by StarPU for one time step,
    including output of the initial and final solutions. The solution \varu~is first copied 
    subdomain per subdomain into a global buffer \dFO, after which this buffer is written in a file 
    by a single task. Since two different descriptors of the buffer are needed, a \texttt{switch} 
    task is launched between both actions.}
\end{figure}

\begin{figure}[ht!!]
  \centering
  \includegraphics[width=1.0\tw]{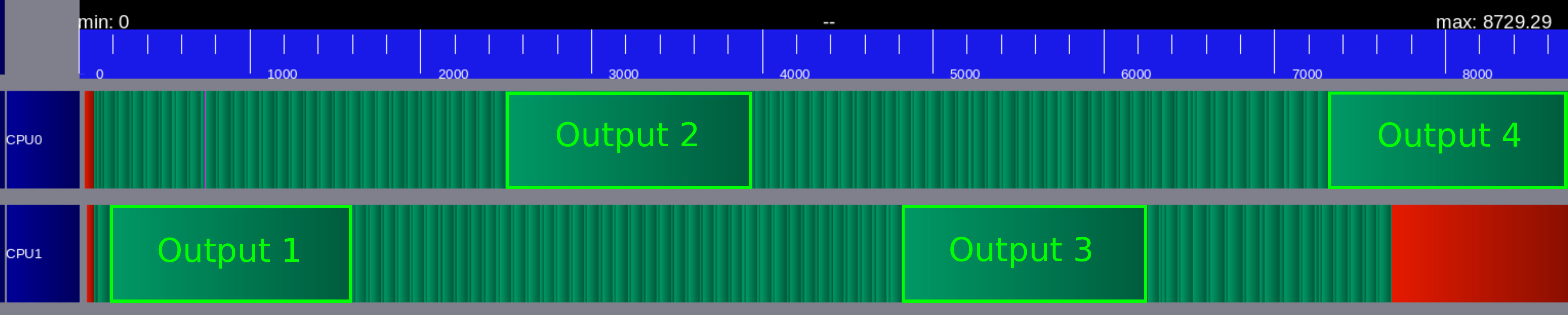}
  \caption{\label{fig:GanttWithOutput} Gantt chart for $60$ time iterations of the finite volume scheme on two cores with an output every $20$ iterations. }
\end{figure}

Eventually, when the solution needs to be written on the disk, the subdomain-wise descriptor is 
activated and each subdomain copies its local solution 
\uLoc~to the appropriate \dFO~sub-data-handle (\texttt{gatherForOutput} task). 
Next, the description of the \dFO~buffer is switched to its global descriptor
and the write-to-disk single task can be executed (\texttt{outputToDisk} task). 
Once the writing is finished, the description of the memory buffer 
is switched back to the partitioned sub-data-handles. 
Figure \ref{fig:TaskDiagramWithOutput} illustrates how the simple task diagram for the numerics 
shown in Figure \ref{fig:TaskDiagramNoOutput} is updated for outputs. 
In Figure \ref{fig:GanttWithOutput}, we plot the Gantt chart obtained for $60$ time iterations of the finite volume scheme on two cores with an output every $20$ iterations. Green color indicates working time and red color symbolizes sleeping time. 
This asynchronous treatment of the outputs allows to perform outputs without blocking one core for the outputs 
and without inducing sleeping time for the other cores.

\section{Results}
\label{sec:Results}

In this last section, we study the performance of our task-driven implementation on 
test cases solving the two-dimensional Euler equations over the periodic domain 
$\left(\mathbb{R}\bigg/\mathbb{Z}\right)^2$. 
More precisely, the state vector and the associated normal flux are given by 
\beq{eq:Euler}
  \W = \left(\rho,\rho u, \rho v, \rho E\right)^t, \quad
  \F.\n(\W) = \left(\rho \un, \,\rho u \un + p n_x, \,\rho v \un + p n_y, \,\rho \un H\right)^t,
\eeq
where $E$ is the \textit{total energy},
$p$ is the \textit{pressure} and $H = E+p/\rho$ is the \textit{total enthalpy}. 
In the following, we assume that $p$ satisfies a perfect gas pressure law  $p= (\gamma-1)\rho(E -\frac{u^2+v^2}{2})$. 
The source term is set to zero: $S(W) = 0$. 
This system of equations \eqref{eq:Euler} is known to be hyperbolic with 
characteristic velocities given by $\un -c$, $\un$, $\un$ and $\un+c$, 
$c$ being the \textit{sound velocity} given by $c^2 = \gamma p /\rho$. 

\subsection{Test cases}

On this set of equations, we use two classical test cases,
namely a linear advection of a density perturbation and an isentropic vortex. 
Both have the advantage to present an analytical solution.

\subsubsection{Localized cosine perturbation on $\rho$}
\label{sec:localized_cosine}

First, we consider an arbitrary perturbation of the density field $\rho(x,y)$ which is simply advected 
on a constant field $\bar{u}$, $\bar{v}$, $\bar{p}$. We choose: 
\beq{eq:LocalizedCosine}
  \begin{array}{c}
  r(x,y) = \sqrt{(x-0.5)^2+(y-0.5)^2}, \quad 
  \rho(x,y) = \rho\left(r(x,y)\right) = 1+(r(x,y)<0.25)*\cos\left(4\pi r(x,y)\right), \\
  \\ 
  \barU = \barV = 1, \quad \text{ and } \quad 
  p = 1/\gamma, 
  \end{array}
\eeq
so that the sound velocity is everywhere smaller than one.

At any computational time $t$, the exact solution is given by
\beq{eq:LCAnalytic}
  r(x,y,t) = \sqrt{(x-\barU t-\floor{x-\barU t}-0.5)^2 + 
                   (y-\barU t-\floor{y-\barU t}-0.5)^2},
  \quad \rho(x,y,t) = \rho(r(x,y,t)),
\eeq
where the operator $\floor{\cdot}$ stands for the floor function, 
in order to take the periodicity into account.

\subsubsection{Isentropic vortex}

Next, we consider a slightly more complex test case which couples all four equations but 
still provides a quasi-analytical solution. The "\textit{quasi}" prefix is explained later on. 
The initial solution is \textit{isentropic} and its velocity field is a superposition of 
a constant and a divergence-free fields:
\beq{eq:isentropy}
  s = \frac{p}{\rho^{\gamma}} = \frac{1}{\gamma}, \quad 
  \u(x,y) = \boldsymbol{\bar{u}}+\boldsymbol{\tilde{u}}(x,y), \; s.t. \;
  \nabla\cdot\boldsymbol{\tilde{u}} = 0.
\eeq
From these two last statements, it is obvious that the entropy and the density are simply advected: 
\beq{eq:Advection}
 \Dt s = 0 \quad \text{ and } \quad \Dt \rho = 0,
\eeq 
$\Dt. = \partial_t. + \u\cdot\nabla.$ being the \textit{material derivative}.

Now, one would like the velocity field to be simply advected by its constant part 
$\boldsymbol{\bar{u}}$, and this comes 
if and only if:
\beq{eq:VelocityAdvection}
  \left(\boldsymbol{\tilde{u}}\cdot\nabla\right) \u + \frac{\nabla p}{\rho} = 0.
\eeq
This is verified on $\mathbb{R}^2$ by an isentropic rotating Gaussian field: 
\beq{eq:RotatingGaussian}
  \omega(r(x,y)) = \bar{\omega}\exp\left(-\frac{r^2(x,y)}{2R^2}\right), \quad
  \boldsymbol{\tilde{u}}(x,y) = \left(-\frac{y}{R} \omega(r), \frac{x}{R}\omega(r)\right)^t, \quad
  \rho(r) = \left(1-\frac{\gamma-1}{2}\omega^2(r)\right)^{\frac{1}{\gamma-1}},
\eeq
where $r$ is the same radius function as in \eqref{eq:LocalizedCosine}, $R$ is a characteristic
radius of the central perturbation and $\bar{\omega}$ is the vortex intensity. 
The numerical values of our setup are: 
\beq{eq:SetUp}
 \bar{\omega} = 1,\quad R=0.1,\quad \barU=\barV=1,\quad \gamma = 1.4.
\eeq
Since the density field is radial, the divergence-free field does not modify it and the 
whole initial solution is just advected by the constant velocity field 
$\boldsymbol{\bar{u}} = \left(\barU,\barV\right)^t$, so that the analytical relation 
\eqref{eq:LCAnalytic} almost still stands true. In fact, the whole previous development is 
true over $\mathbb{R}^2$ but the connection of the velocity field 
at the boundaries of our periodic domain $\Omega$ is not continuous. Nonetheless, the jump 
exponentially decreases when the size of the domain increases, which comes exactly to the same 
as decreasing $R$. Then, for a given final time $T$, we can choose $R$ such that the 
error of our approximated analytical solution is negligible compared to the numerical error, 
so that the latter can be measured, \cite{Nasa15,shuVortex}.

\subsubsection{Validation by mesh convergence study}

\begin{figure}
	\begin{center}
		\subfloat[\small Localized cosine. Straight lines are estimated slopes of the order of $0.95$.]{
			\begin{minipage}[c]{0.5\textwidth}
				\includegraphics[width=0.6\tw,angle=-90,clip,trim=40 0 0 0]{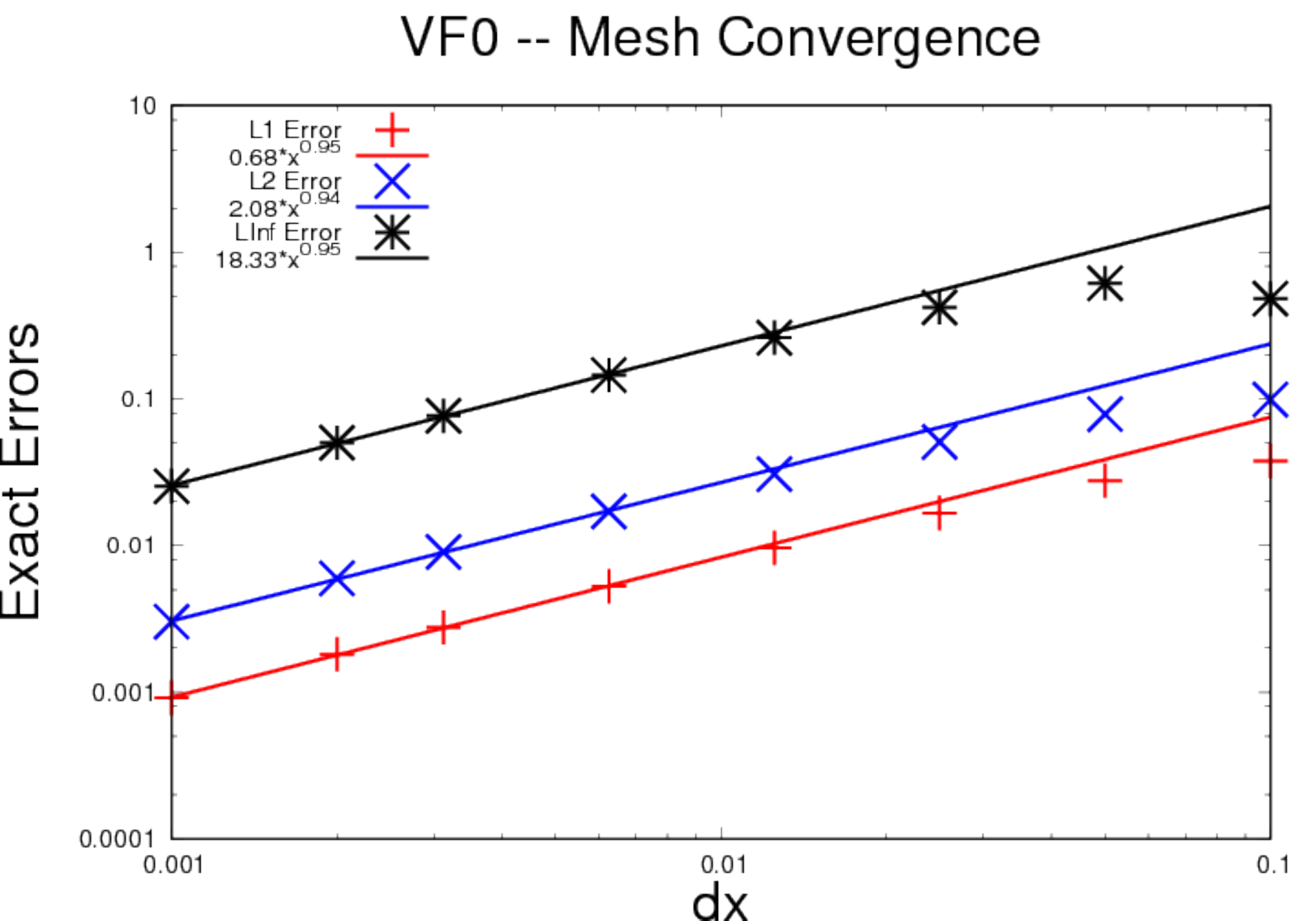}
			\end{minipage}
    }
		\subfloat[\small Isentropic vortex. Straight lines are estimated slopes of the order of $0.95$.]{
			\begin{minipage}[c]{0.5\textwidth}
				\includegraphics[width=0.6\tw,angle=-90,clip,trim=40 0 0 0]{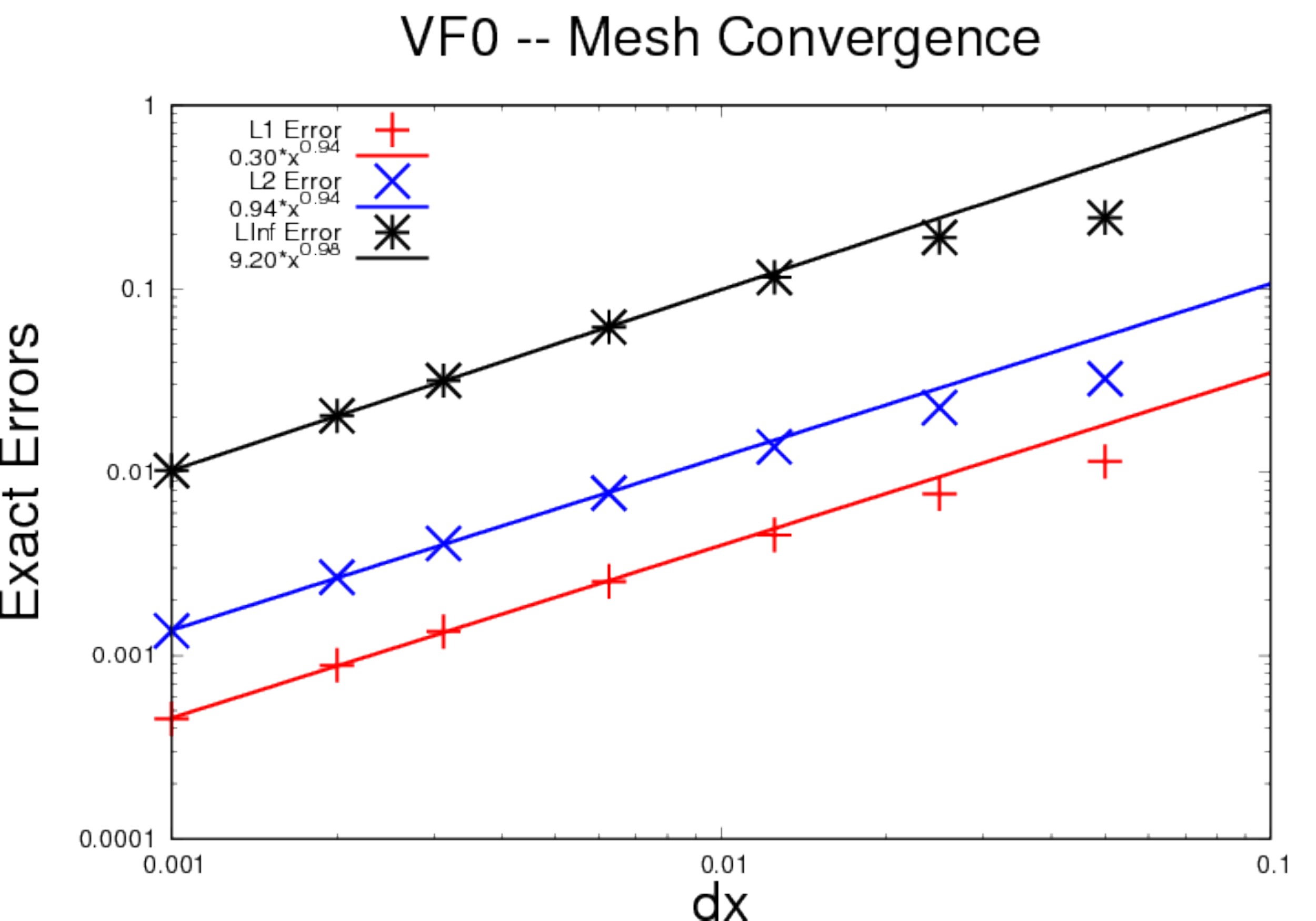}
			\end{minipage}}
		\caption{Mesh convergence study for both test cases. Errors are $L^1$ (red), $L^2$ (blue) }
		\label{fig:Convergence}
	\end{center}
\end{figure}

In Figure \ref{fig:Convergence} we show the convergence results obtained on both test cases 
described above. This validates the implementation of our numerical method since it is of 
order $1$, as expected. 

\subsection{Parallel efficiency}
\begin{figure}
\begin{center}
		\subfloat[\small 2 dodeca-core Haswell Intel Xeon E5-2680.]{
			\begin{minipage}[c]{0.5\textwidth}
				\includegraphics[width=1.0\tw]{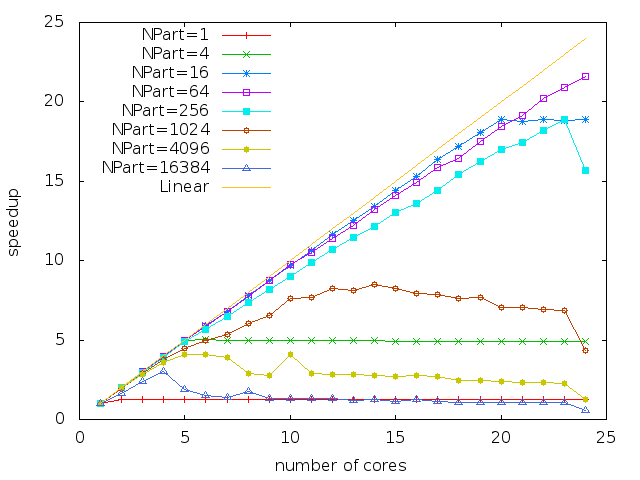}
			\end{minipage}
      \label{fig:SSE-A}
    }
		\subfloat[\small Xeon Phi KNL.]{
			\begin{minipage}[c]{0.5\textwidth}
				\includegraphics[width=1.0\tw]{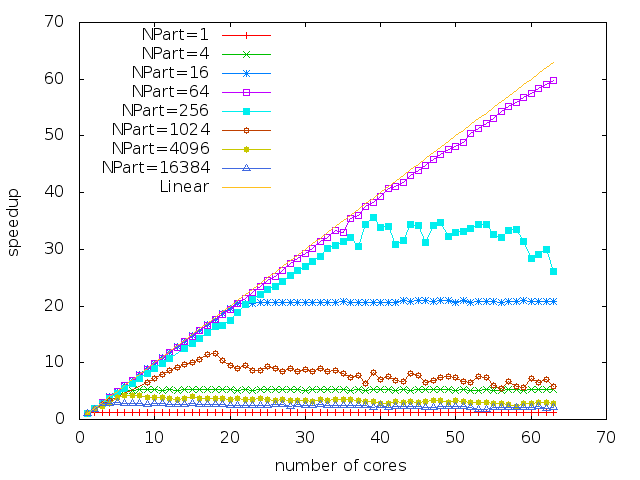}
			\end{minipage}
      \label{fig:SSE-B}
    }
		\caption{Strong scaling with the \texttt{eager} scheduler: scalability results obtained on two different architectures with a number of subdomain $\NPart$ varying from $4$ to $16384$.}
		\label{fig:StrongScalingEager}
	\end{center}
\end{figure}

We perform a strong scaling study. It means that we keep the same problem size and we increase the number of cores. In Figure \ref{fig:StrongScalingEager}, we consider a mesh of $1024\times 1024$ cells subdivided in $\NPart=1, 4, ..., 16384$ sub-domains and we perform a scalability study on two different architectures: a 2 dodeca-core Haswell Intel Xeon E5-2680 and a Xeon Phi KNL. Recall that the number of tasks is proportional to the number of sub-domains $\NPart$. To have a correct scaling, we need enough tasks. $\NPart$ should be at least of the order of the number of cores. However, if we increase the number of tasks $\NPart$ a lot, the size (time duration) of each task becomes to small compared to the task overhead (see section \ref{ssec:Overhead}) and the scaling starts to saturate. 
To sum up, in order scale correctly on many cores, 
we need a lot of tasks of long duration (compared to the tasks management overhead) 
and therefore, we need big meshes.

In Figure \ref{fig:StrongScalingEager}, we distinguish three types of curves. 
Those for which we do not provide enough tasks: 
$\NPart = 1,4,16$ in \ref{fig:SSE-A} and \ref{fig:SSE-B}.
Those which scale correctly on all the cores: 
$\NPart = 64,256$ in \ref{fig:SSE-A}, 
$\NPart = 64$ in \ref{fig:SSE-B}.
And finally, those who start to saturate due to task overhead: 
$\NPart = 1024,4096,16384$ in \ref{fig:SSE-A} and the same partitions including 
$\NPart = 256$ in \ref{fig:SSE-B}.

\subsection{GPU implementation and performance}

\subsubsection{GPU implementation}

In order to launch part of the computation on GPUs, one just has to write a CUDA or OpenCL 
version of the kernels, see paragraph \ref{sec:StarPUGlossary}. We choose to use CUDA. 
Then, we let the scheduler choose between a CPU device or a GPU device, for each task.
It is not necessary to have a CUDA implementation of each kernel, 
but for performance reason (to minimize the communication between the CPU and the GPU) 
we need at least to write a CUDA version of the following kernels: 
\texttt{checkTimeStep}, \texttt{copyOverlaps}, \texttt{internalResidual}, 
\texttt{borderResiduals} and \texttt{update}, see paragraph \ref{sec:Tasks_description}. 
Let us detail the CUDA implementation of these kernels: 
\begin{itemize}
\item \texttt{copyOverlaps}: we associate a virtual processor for each variable of each cell and we copy \texttt{uLoc} into \texttt{olvpE}, \texttt{olvpN}, \texttt{olvpW} or \texttt{olvpS}. For \texttt{olvpN} and \texttt{olvpS}, since two successive processing elements access two neighboring memories (in global memory), it allows to achieve optimal memory bandwidth (coalescing access). This is not the case for \texttt{olvpE} and \texttt{olvpW}.
\item \texttt{update}: we associate a virtual processor for each variable of each cell and we add \texttt{RHS} to \texttt{u}, see \eqref{eq:update}. The memory accesses are also coalescent.
\item \texttt{checkTimeStep}: we associate a virtual processor for each cell. The checking is a little bit different from the CPU implementation. Indeed, computing the largest characteristic speed is quite difficult in parallel. Then, we compute a local time step
\[
(\Delta t)_{i,j}=\frac{\min(\Delta x,\Delta y)}{\underset{p\in\llbracket 1,m\rrbracket}{\max}\vert \lambda_p(W_{i,j})\vert}
\]
and we check if $(\Delta t)_{i,j}$ is greater than the constant time step $\Delta t$.
\item \texttt{internalResidual} and \texttt{borderResidual}: we associate a virtual processor for each cell. We compute the numerical flux on each edge of this cell and we add the result to the right hand side, \texttt{RHS}, of this cell. 
In the GPU implementation, we compute the fluxes twice: 
the two vitual processors associated to cells $C_{i,j}$ and $C_{i+1,j}$ 
compute the same flux $\FStar_{i+\frac{1}{2},j}$. 
Indeed, computing the flux once at each edge and balancing it on the two neighboring cells 
would induce concurrent memory access. For example, the fluxes computed at edges 
$\Gamma_{i-\frac{1}{2},j}$ and $\Gamma_{i+\frac{1}{2},j}$ could be added simultaneously to 
$\texttt{RHS}_{i,j}$.
However, since computation is fast on GPU, even if we compute all the fluxes twice, 
we still observe good performance.
\end{itemize}

\subsubsection{Performance model}
\label{subsec:PerfModel}
To reach good scheduling performances on heterogeneous architectures 
(with CPU and GPU for example), 
StarPU needs to be able to estimate in advance the duration of a task. 
This is done by associating each codelet with a performance model. 
This performance model provides for each codelet and for a certain range of data sizes, 
an expected average and standard deviation of the task execution times. 

\begin{figure}
	\begin{center}
		\subfloat[\small Codelet \texttt{checkTimeStep}.]{
			\begin{minipage}[c]{0.5\textwidth}
				\includegraphics[width=0.8\tw]{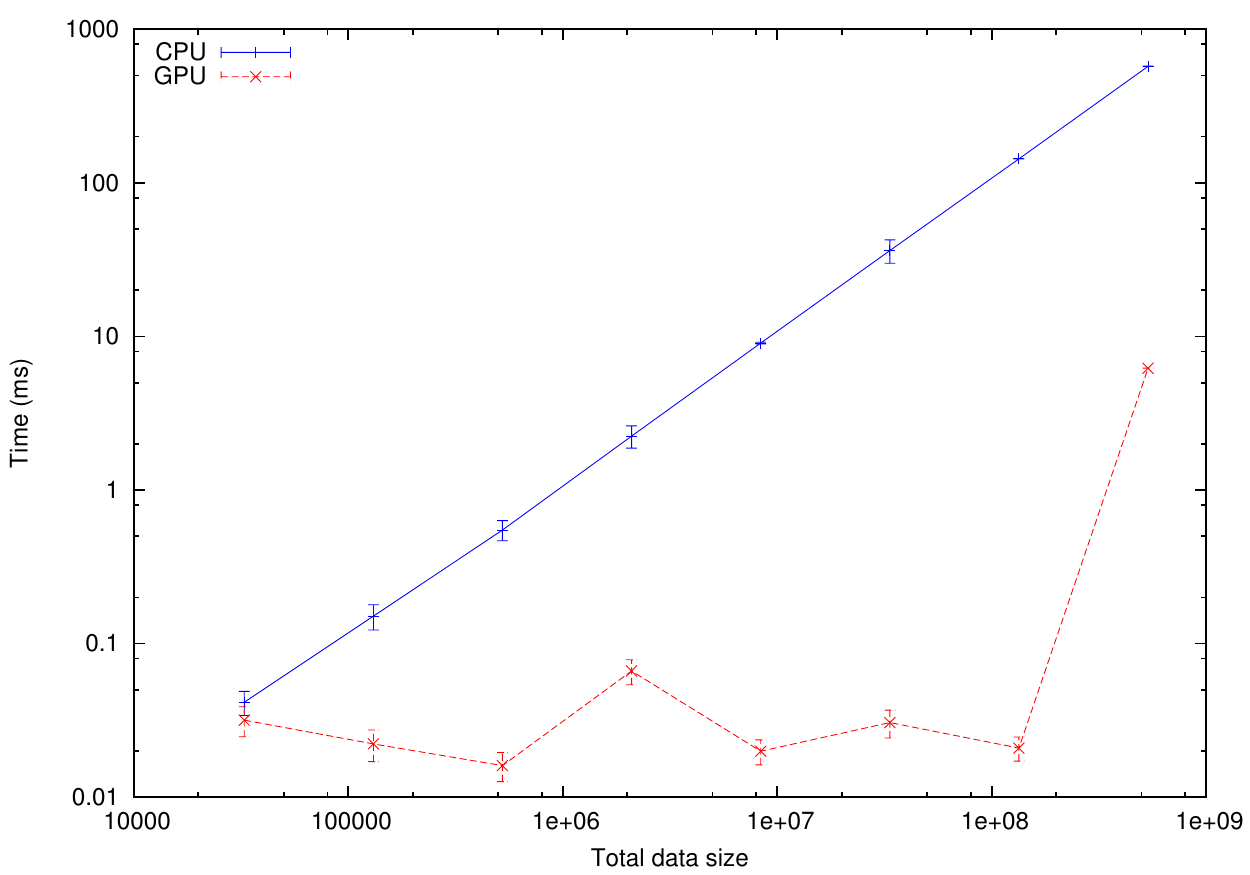}
			\end{minipage}
    }
		\subfloat[\small Codelet \texttt{internalResidual}.]{
			\begin{minipage}[c]{0.5\textwidth}
				\includegraphics[width=0.8\tw]{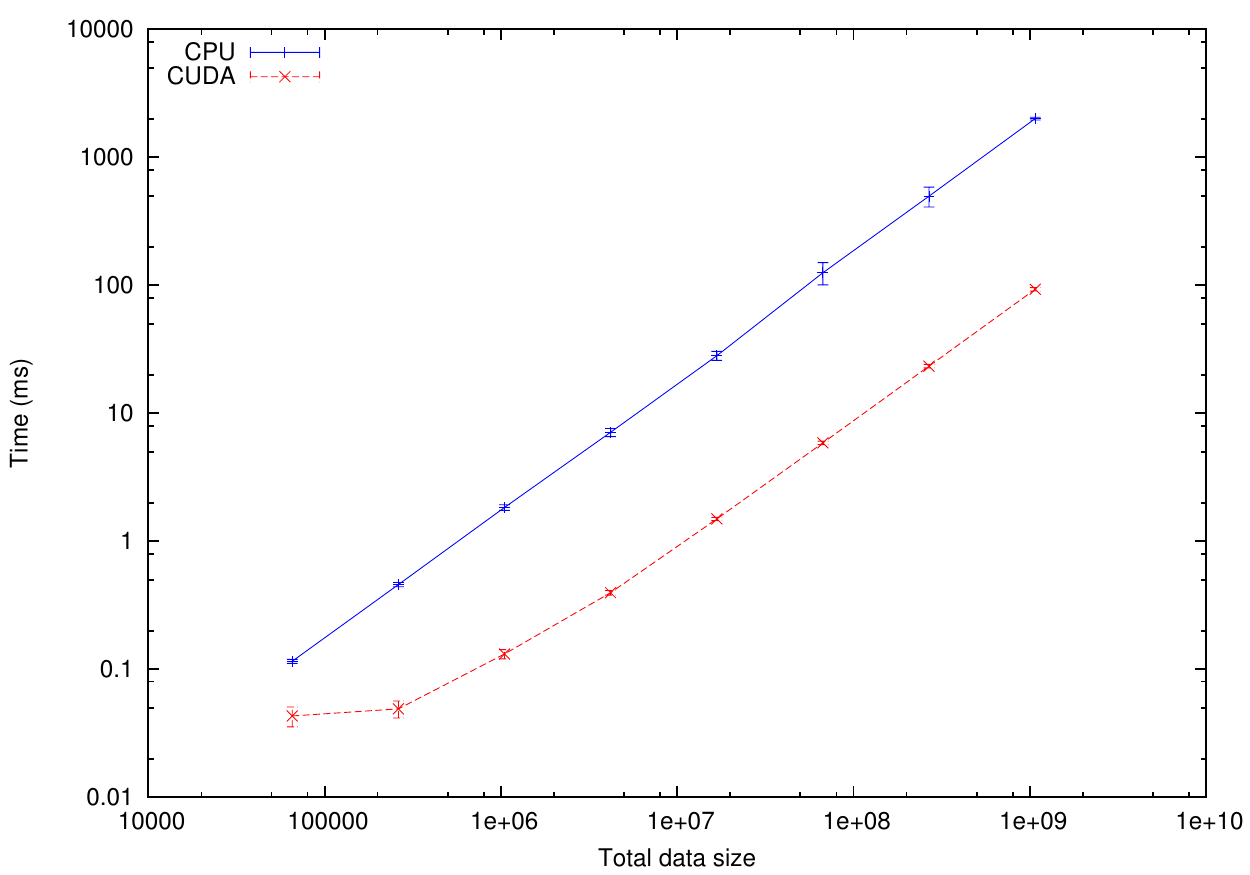}
			\end{minipage}}\\
		\subfloat[\small Codelet \texttt{borderResidual}.]{
			\begin{minipage}[c]{0.5\textwidth}
				\includegraphics[width=0.8\tw]{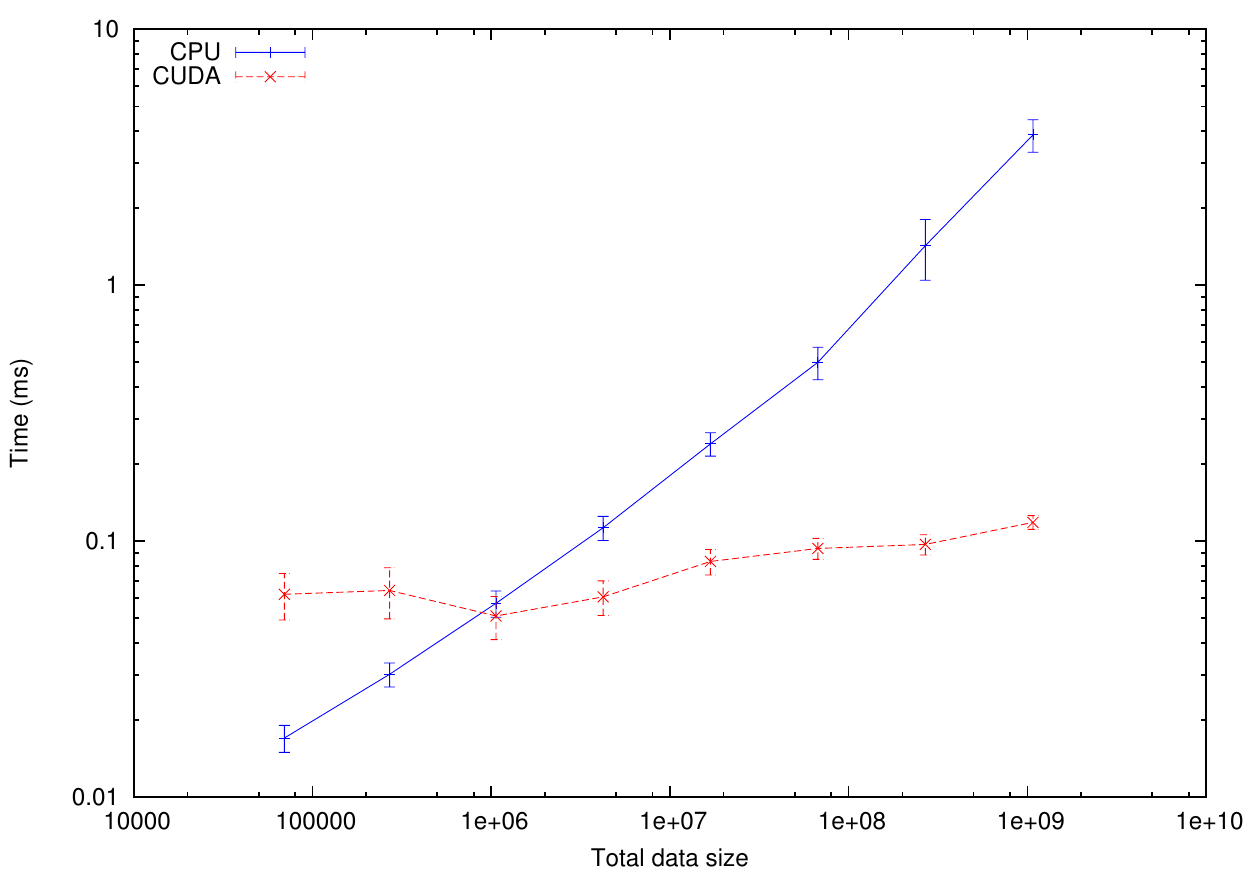}
			\end{minipage}
    }
		\subfloat[\small Codelet \texttt{update}.]{
			\begin{minipage}[c]{0.5\textwidth}
				\includegraphics[width=0.8\tw]{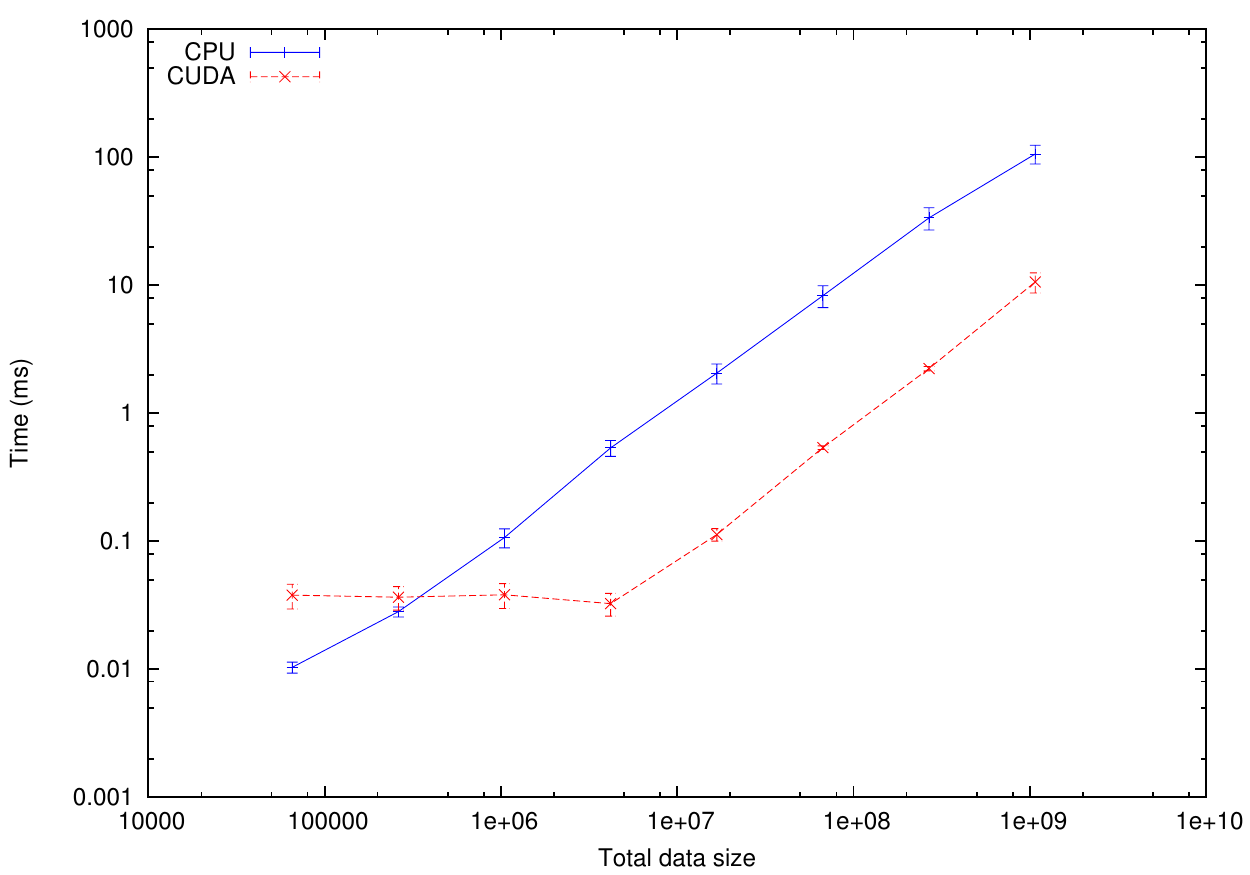}
			\end{minipage}}
		\caption{Average of the execution times on CPU and GPU of the tasks associated to a specified codelet as a function of the data size of the tasks.}
		\label{fig:Performance_1}
	\end{center}
\end{figure}

\begin{figure}
	\begin{center}
		\subfloat[\small Codelet \texttt{copyOverlap South}.]{
			\begin{minipage}[c]{0.5\textwidth}
				\includegraphics[width=0.8\tw]{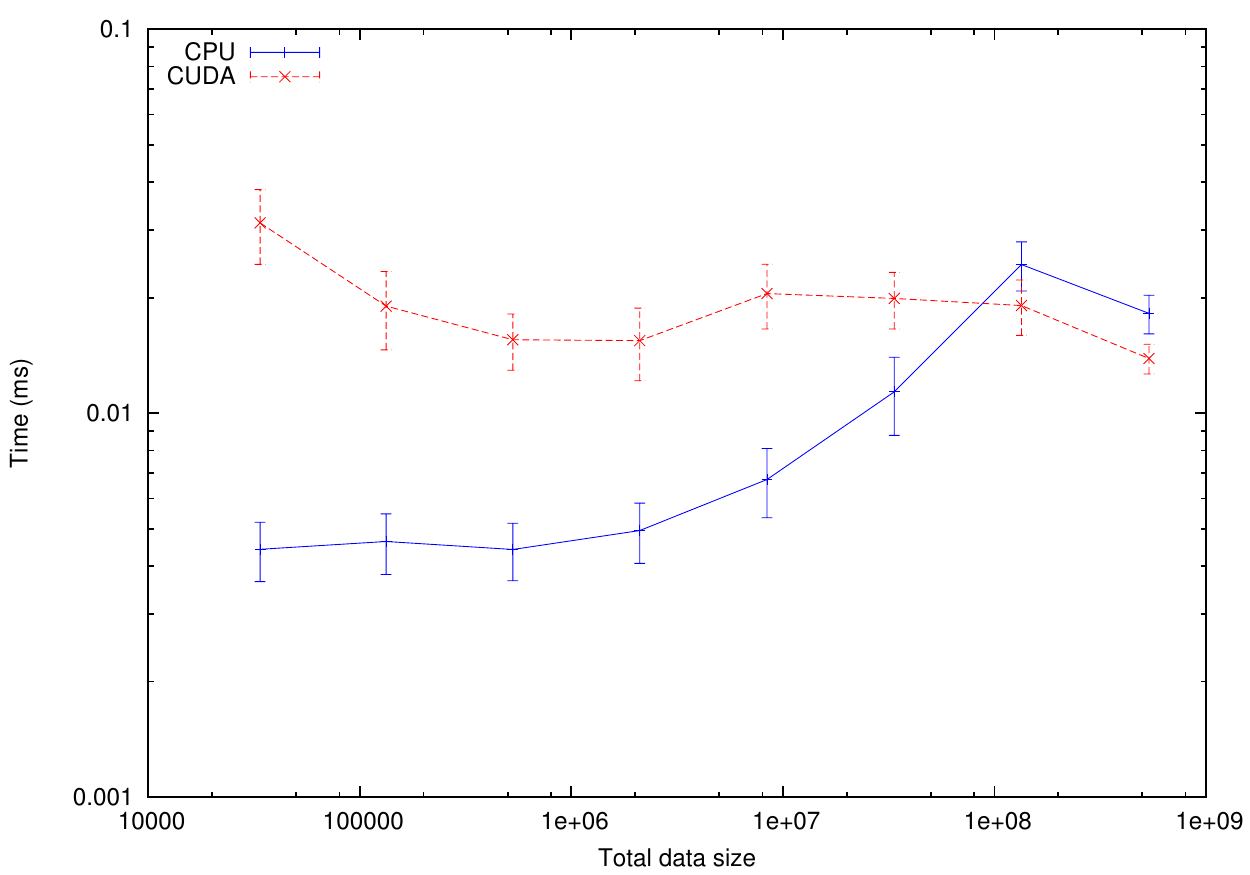}
			\end{minipage}
    }
		\subfloat[\small Codelet \texttt{copyOverlap North}.]{
			\begin{minipage}[c]{0.5\textwidth}
				\includegraphics[width=0.8\tw]{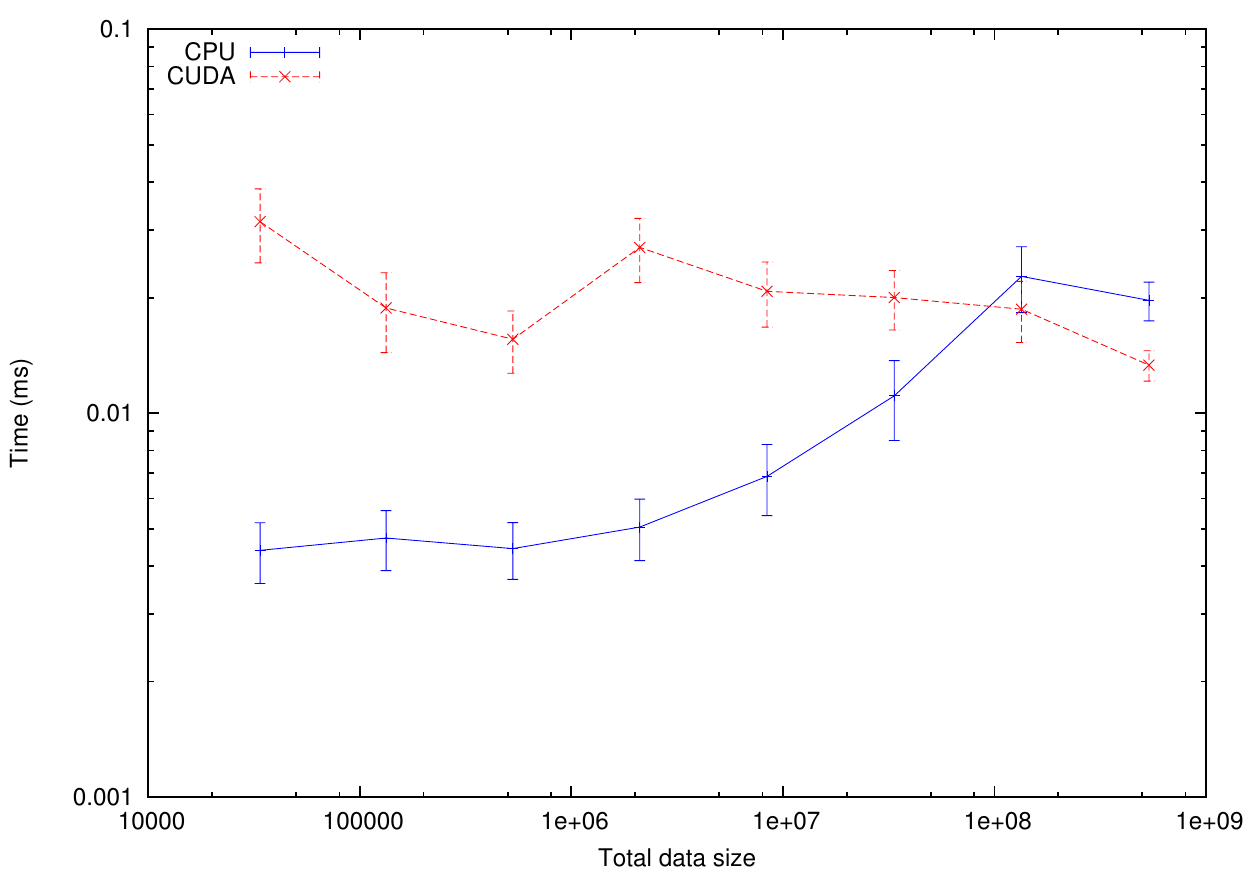}
			\end{minipage}}\\
		\subfloat[\small Codelet \texttt{copyOverlap Est}.]{
			\begin{minipage}[c]{0.5\textwidth}
				\includegraphics[width=0.8\tw]{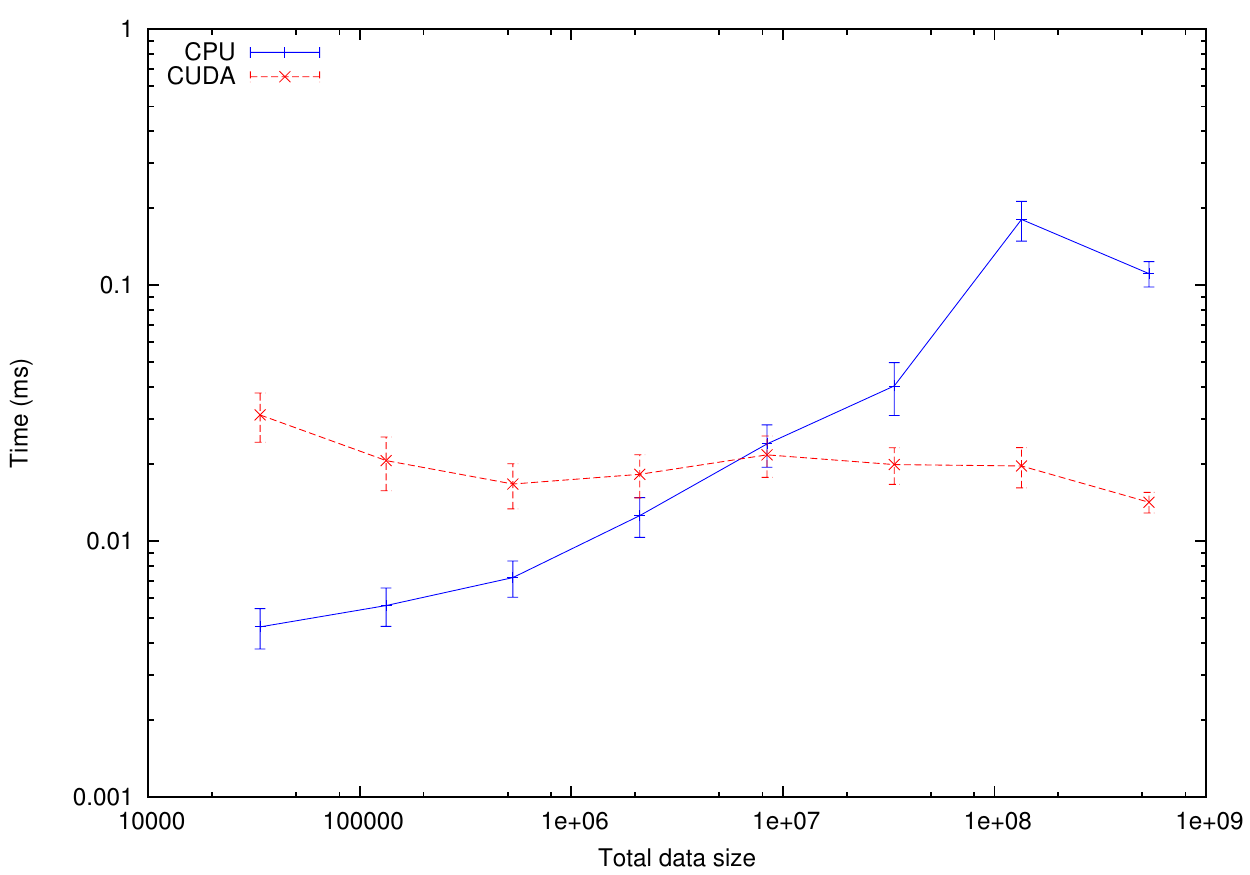}
			\end{minipage}
    }
		\subfloat[\small Codelet \texttt{copyOverlap West}.]{
			\begin{minipage}[c]{0.5\textwidth}
				\includegraphics[width=0.8\tw]{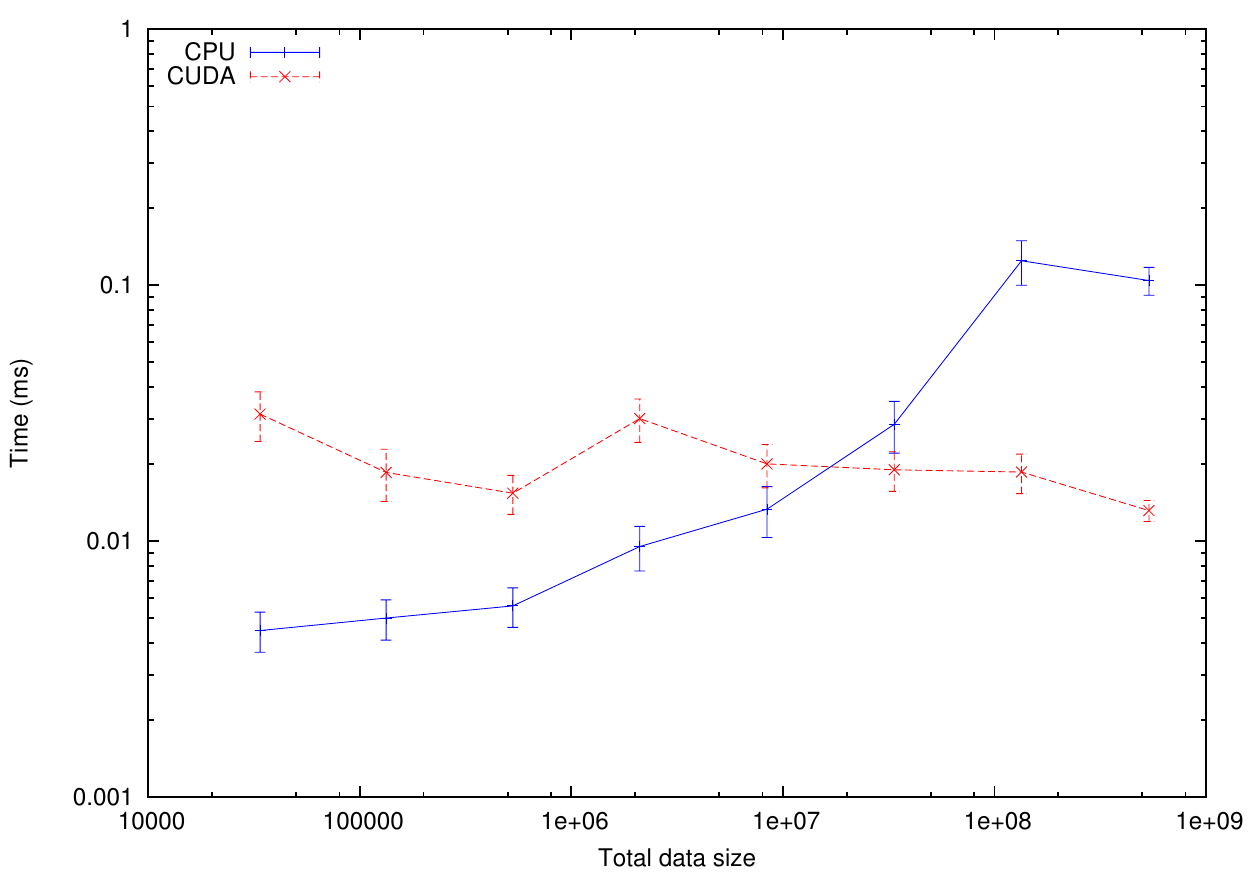}
			\end{minipage}}
			
		\caption{Average of the execution times on CPU and GPU of the tasks associated to a specified codelet as a function of the data size of the tasks.}
		\label{fig:Performance_2}
	\end{center}
\end{figure}

We use this tool to compare the execution time  of different tasks 
on a CPU Haswell Intel Xeon E5-2680 and on a GPU Nvidia K40-M. 
We consider the numerical test of Section \ref{sec:localized_cosine}. The  computational domain is divided into $8192\times 8192$ cells. In Figures \ref{fig:Performance_1}-\ref{fig:Performance_2}, we plot the average execution times on CPU and GPU of the tasks associated to a specified codelet as a function of the tasks data size. To decrease the tasks data size, we increase the number of sub-domains from $2\times 2$ up to $256\times 256$. 
The maximum size of the tasks is fixed by the memory of the GPU: $10$~GB in our case.

From Figures \ref{fig:Performance_1}-\ref{fig:Performance_2}, we immediately notice that 
computationally expensive tasks, such as \texttt{checkTimeStep}, 
\texttt{internalResidual} and \texttt{update}, 
are greatly accelerated on GPU. Observed speedups are between $2.6$ and $24$ for 
\texttt{internalResidual} and between $0.33$ and $20$ for \texttt{update}.
However, tasks involving more data transfer than useful calculations, such as every task 
associated with subdomains boundaries (\texttt{copyOverlap} and \texttt{borderResidual}), 
present no reason to be executed on a GPU in the range of data size commonly used. 
This is where the scheduler starts playing an important role: the overall computation will 
accelerate if part of its tasks are executed on the GPUs, but certainly not all of them. 

\subsubsection{Graph of activities with CPUs and GPUs}

\begin{figure}
	\begin{center}
		\subfloat[\small \texttt{eager} scheduler.]{
			\begin{minipage}[c]{0.5\textwidth}
				\includegraphics[width=0.8\tw]{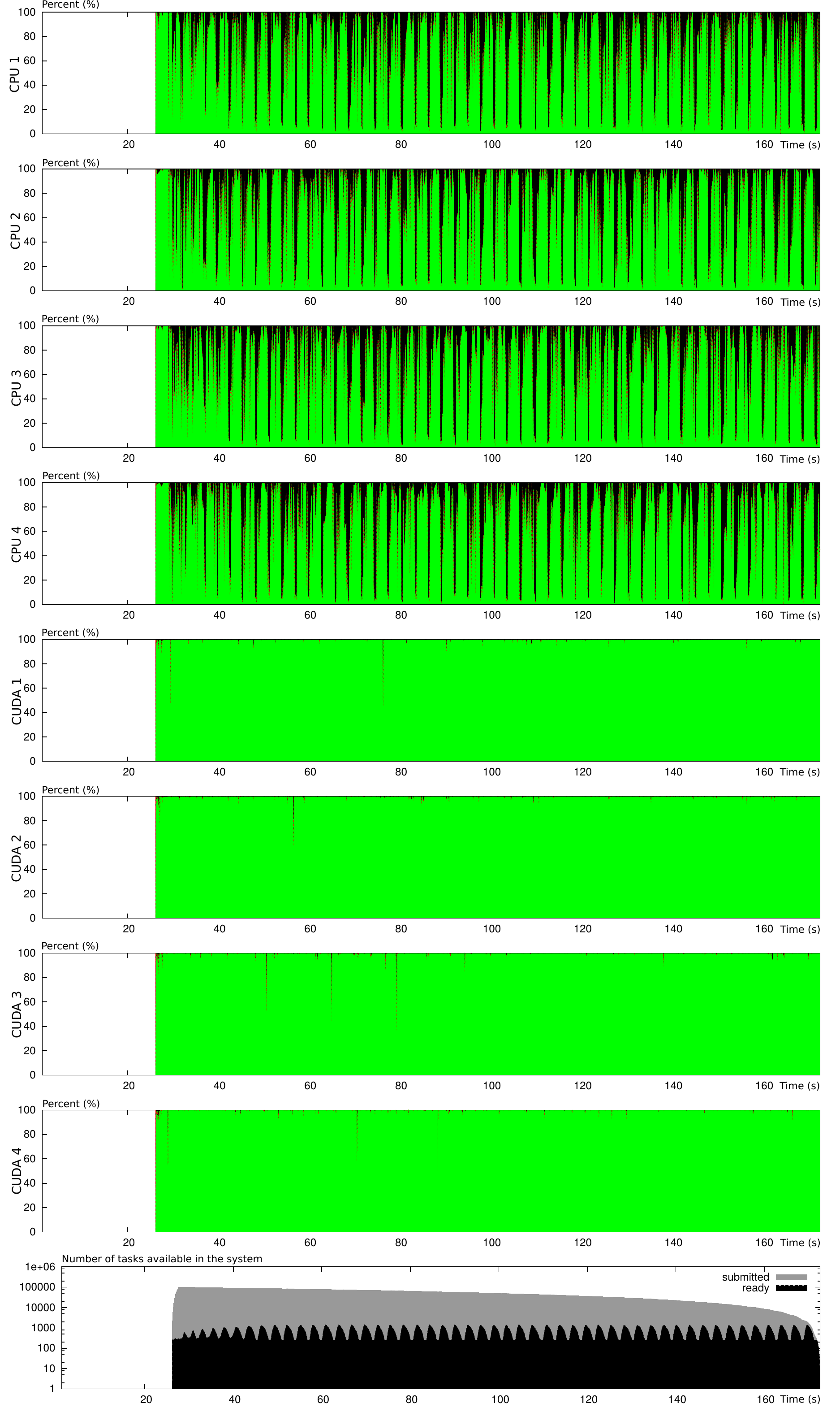}
			\end{minipage}
    }
		\subfloat[\small \texttt{dmda} scheduler.]{
			\begin{minipage}[c]{0.5\textwidth}
				\includegraphics[width=0.8\tw]{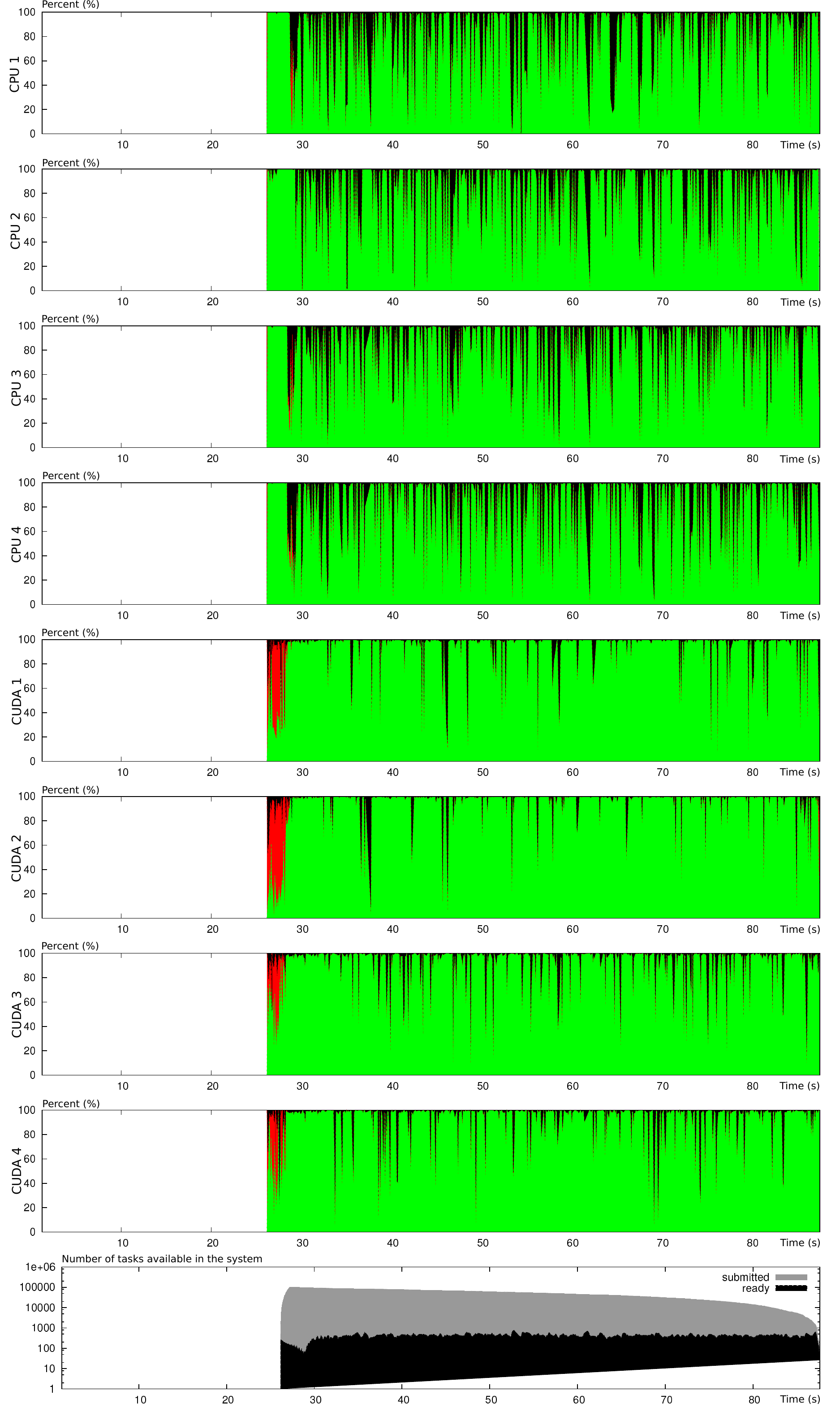}
			\end{minipage}
    }
			
		\caption{Comparison of the activities obtained with the \texttt{eager} and the \texttt{dmda} schedulers during the computing time. }
		\label{fig:Activities}
	\end{center}
\end{figure}

With StarPU, there are different tools to study the activities of each worker. We look here at 
the graph of activities: it shows the activity of each worker and the evolution of the number of tasks available on the system with respect to the execution time.
In Figure \ref{fig:Activities}, we plot the graph of activities obtained with a run of $50$ iterations of the finite volume scheme on a $16 384\times 16 384$ cells domain, 
subdivided in $16\times 16$ sub-domains, 
on a heterogeneous architecture of $4$ CPUs and $4$ GPUs. 
We consider the work done by two different schedulers: \texttt{eager} and \texttt{dmda}. 
Green sections indicate time spent on the kernels execution. Red sections indicate 
the proportion of time spent in StarPU.
Black sections indicate sleeping time. 
With both schedulers, our graphs are essentially green, which means that all our workers have enough work: the granularity of the tasks is good and thus the parallelism is also good. 
We have a lot of tasks ($16\times 16$ sub-domains) and each sub-domain stays big 
(more than four millions of unknowns). 
However, even if we observe only little waiting time on both graphs, 
we remark that the execution time is more than two times smaller with the \texttt{dmda} scheduler than with the \texttt{eager} one ($61$~s versus $146$~s). 
In fact, the \texttt{dmda} scheduler takes the performance models into account and 
therefore executes every task on its optimized device.
These graphs are very useful to know the activities of each worker but do not give any information about the tasks executed on each worker. 
To obtain such information one would need to plot a Gantt chart, which is an other performance
tool provided by StarPU. This is now the main content of the next section which should 
be considered as a concluding and opening section. 

\section{Opening: toward HPC of complex multiphase flows on large scale heterogeneous clusters}

As we have seen during the last result section, the performance of the task driven 
implementation strongly depends on the work load of each task. In a nutshell, 
we need a large number of big tasks. This is simply achieved by increasing 
the local workflow, which could be achieved by increasing the number \nVar~of equations,
by increasing the number \nDoF~of degrees of freedom per cell, or by adding some 
local work by mean, for example, of a source term. We start by considering the latter case.

\subsection{A model for the dynamics of an evaporating disperse phase} 

Without entering some complex explanation, we consider the following hyperbolic system 
of conservation laws, \cite{EssadkiSiam16}:
  \begin{equation}\label{eq:Essadki}
    \left\{
      \begin{array}{ccccccl}
        \partial_t m_{0/2}
        &+& 
        \partial_{\xv}\cdot\left(m_{0/2} \u\right)
        &=&
        -\KEvap \; n(t,\xv,S=0),
        & & 
        \\
        \partial_t m_{1/2}
        &+& 
        \partial_{\xv}\cdot\left(m_{1/2} \u\right)
        &=&
        -\frac{\KEvap}{2}\; m_{-1/2},
        & & 
        \\
        \partial_t m_{2/2} 
        &+& 
        \partial_{\xv}\cdot\left(m_{2/2}\u\right)
        &=&
        -\KEvap \; m_{0/2},
        & & 
        \\
        \partial_tm_{3/2}
        &+& 
        \partial_{\xv}\cdot\left(m_{3/2}\u\right)
        &=&
        -\frac{3\KEvap}{2} \; m_{1/2},
        & & 
        \\
        \partial_t\left(m_{2/2}\u\right)
        &+& 
        \partial_{\xv}\cdot\left(m_{2/2}\u\otimes\u\right)
        &=&
        -\KEvap \; m_{0/2}\u
        &+&
        m_{0/2} \frac{\u_g-\u}{\theta},
      \end{array}
    \right.
  \end{equation}
where $m_{k/2}(t,\xv) = \int_0^1 S^{k/2} n(t,\xv,S) dS$ are the fractional moments of a certain 
size distribution $n(t,\xv,S)$, $S$ indexing the size, evaporating at rate $\KEvap$ and $\u(t,\xv)$ 
is its velocity field which relaxes toward an underlying gas field $\u_g$ at time rate $\theta S$, 
thanks to the bottom right term. 

This system has now $\nVar=6$ equations. The left hand part is a simple linear transport along the velocity 
field $\u(t,\xv)$. The right hand side is made of two source terms: an evaportation term acting on all the moments 
and a drag term acting on the velocity components only, modeling the traction of the spray by an underlying gas field. 
As discussed in \ref{sssec:Source}, the system terms are split so that the resolution of \eqref{eq:Essadki} 
boils down to 
\beq{eq:Splitting}
  \partial_t \M + \partial_{\xv}.\F(\M) =0, 
  \qquad
  \text{d}_t\M = -\KEvap\left[n(t,\xv,S)\right]
  \quad \text{ and } \quad
  \text{d}_t(m_{2/2} \u) = m_{0/2} \frac{\u_g-\u}{\theta}.
\eeq

\subsection{Integration strategy}

Following what has been said in previous sections, our numerical procedure is rather simply adapted to this new 
system. The only key point lies in the evaporation term which requires the reconstruction of the entire size distribution 
$n$ from the moments $m_{k/2},\;k=0,\dots,3$ by an entropy maximization procedure, see \cite{EssadkiSiam16} for more details.
This allows to estimate the unknown quantities $n(t,\xv,S=0)$ and $m_{-1/2}$.
Our first order numerical procedure being convex state preserving, the set of moments is everywhere realizable and  
this local reconstruction is always possible in every cell. However, the local reconstruction of $n$ 
involves the computation of exponential of matrices which are rather costful, especially when compared to the simplicity 
of the other terms. 

\begin{figure}
	\begin{center}
		\subfloat[\small $0$ GPU. Eager scheduler. Execution time = 32.930s.]{
				\includegraphics[width=0.99\tw]{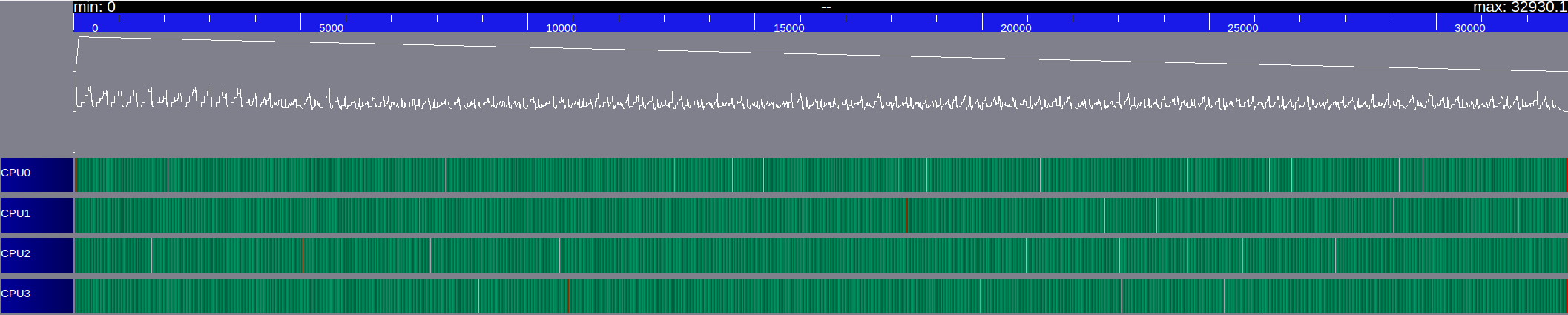}
        \label{fig:0GPU}
    }\\
		\subfloat[\small $1$ GPU. Eager scheduler. Execution time = 12.465s.]{
				\includegraphics[width=0.99\tw]{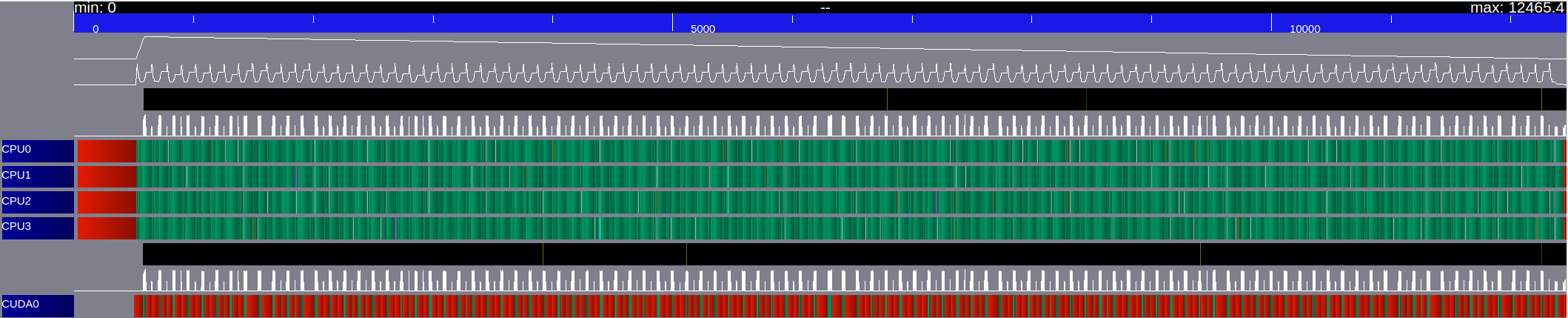}
        \label{fig:1GPU_eager}
    }\\
		\subfloat[\small $1$ GPU. Source terms forced on GPU. Execution time = 5.810s.]{
				\includegraphics[width=0.99\tw]{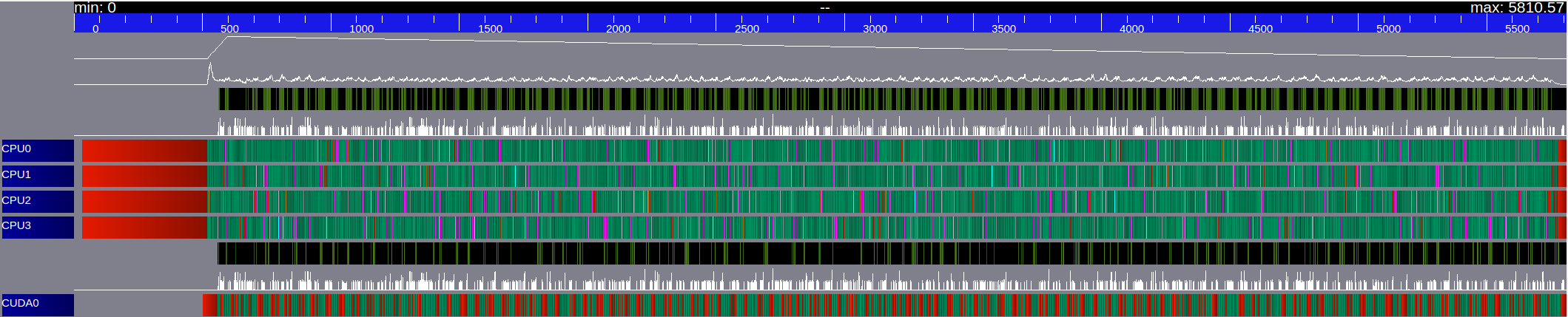}
        \label{fig:1GPU_dmda}
    }
		\caption{Gantt chart of the tasks distribution for different configuration. TOP: no GPU is activated. MIDDLE: one GPU is 
      active and tasks are distributed thanks to the \texttt{eager} scheduler. 
      BOTTOM: one GPU and \texttt{dmda} scheduler.}
		\label{fig:WithSource}
	\end{center}
\end{figure}

This explains why we have tried to pass this reconstruction procedure on GPUs and compared the computational times 
when activating the accelerating device or not and when using one scheduler (\texttt{eager}) or another (\texttt{dmda}). 
The results are presented in Table \ref{tab:WithSource} and in Figure \ref{fig:WithSource}. All simulations hereafter compute $100$ 
time iterations on a $200\times 200$ domain, divided into $2\times2$, $2\times4$, $4\times4$ or $4\times8$ subdomains, 
of an evaporating spray within a Taylor-Green vortices velocity field for the gas, the same as the latest test-case of 
\cite{EssadkiSiam16}. 
In table, \ref{tab:WithSource}, we see first that activating the GPU immediately improves the computational time a lot:
the reconstruction procedure requires fast linear algebra GPUs are very good at. 
Next, the \texttt{eager} scheduler is not smart and does not anticipate the global computational time of the tasks. 
When switching to the \texttt{dmda} scheduler, we gain a little bit more performance by better distributing the tasks 
on the devices: this is illustrated on the Gantt charts in Figure \ref{fig:WithSource}. On the top figure, we look 
at the task scheduling when no GPU is activated. This is the \texttt{eager} scheduler. \texttt{dmda} scheduler 
does only worsen the situation, since each of the four CPUs is permanently loaded with tasks. We do not see it very much 
on this image due to the large number of tasks, but the source term tasks are about four time larger than the others: 80\% 
of the effort is spent on these tasks. However, if we activate the GPU kernel for the source term tasks, we immediately 
reduce the computational time by a factor 2.6, when the \texttt{eager} scheduler is kept, see Figure \ref{fig:1GPU_eager}.
But there, we also see that not all the source term tasks have been given to the GPU, even though it is much faster at 
treating these tasks. This explains why, when turning the \texttt{dmda} scheduler on, Figure \ref{fig:1GPU_dmda}, 
the computational time is again reduced by more than a factor 2: now all the source term tasks are sent to the GPU and 
only the transport part remains on the CPU.

Nonetheless, we also see that some red stripes corresponding to slipping time, remain 
on the CPU line, meaning that one could gain even a little more by activating the computation of the transport tasks 
on the GPU card. 

\begin{table}
  \begin{tabular}{|l|c|c|c|} 
   \hline
   scheduler  & part.      &  \large{0 GPU} & \large{1 GPU} \\
   \hline
   \multirow{4}{*}{\texttt{eager}}
                 & $2\times2$ &     36.06      & 33.12 \\
                 & $2\times4$ &     36.40      & 12.42 \\
                 & $4\times4$ &     37.26      & 10.26 \\
                 & $4\times8$ &     38.87      & 15.37 \\
   \hline
   \multirow{4}{*}{\texttt{dmda}}
                 & $2\times2$ &     69.61      & 13.76 \\
                 & $2\times4$ &     44.76      & 11.46 \\
                 & $4\times4$ &     39.33      & 10.94 \\
                 & $4\times8$ &     39.31      & 12.98 \\
    \hline
  \end{tabular}
   \caption{\label{tab:WithSource} Computational time (in seconds) of the integration of system \eqref{eq:Essadki}
     on a $200\times200$ domain divided into $2\times2$, $2\times4$, $4\times4$ or $4\times8$ subdomains, 
     when activating or not a GPU accelerating unit and using the \texttt{eager} or \texttt{dmda} scheduler.
   }
\end{table}

\subsection{Conclusion and opening}

We acknowledge the last results of this section are poor in term of physical meaning, especially since the numerical method 
is limited at first order and the model integration is not fully assessed. Nonetheless, this paper has stayed technical 
on purpose, in order to explore all the possibilities and restrictions 
of the implementation of a numerical method within the framework of a runtime, namely StarPU. 

We have described and shown experimentally the main features of our code: the numerical procedure described in 
section \ref{sec:FV0} is subdivided into tasks (section \ref{sec:PracticalImplementation}), after a quick 
presentation of the StarPU environnement in section \ref{sec:StarPU}. In section \ref{sec:Results}, we have clearly demonstrated 
that the task distribution is efficient if the tasks are both large and numerous enough, what requires a large enough 
problem at start. Then, we have looked at the task distribution on a heteregeneous architecture and shown 
that all the tasks do not deserve to be executed on the GPU accelerator: only the one with the highest arithmetic intensity 
should be loaded there, while memory consuming tasks should stay on the host node. This demonstrates 
the necessity for an efficient scheduler which is going to make the task distribution decisions on the fly. 

In this last section, we have had the will to go toward a more HPC oriented application. 
Instead of looking at standard Euler equations, 
we have implemented a model for the dynamics of an evoporating spray including intensive source terms computation. 
This test case is very promising for our study. Indeed, we  intend to go to higher order numerical methods, which 
imply an increased number of degrees of freedom per cell and an increased local arithmetic intensity. 
Moreover, this considered model is part of a hierarchy of models which can be enriched at will by involving 
an increasing amount of meaningful variables and associated equations. All of this goes in the right direction, 
which is a larger number of computationally expensive tasks. 

The work at CEMRACS16 within the Hodins team has been fruitful, with many interesting results and a 
common mastering of the task-driven programming for physical applications. As it is now clear, the results 
are promising enough for the collaboration to go on and further studies now need to be lead with higher order 
numerical procedures, which are going to be of the discontinuous Galerkin type. 

\section*{Acknowledgment}

The Hodins team would like to thank the following structures for their support during the 2016 CEMRACS session:
\bitem
  \item Groupe Calcul and Fondation Hadamard for its financial support, 
  \item IFPeN for authorizing M. Essadki to be part of the adventure, 
  \item Aix-Marseille and Plafrim (Bordeaux) computational platforms on which most of the results have been performed, 
  \item ANR Subsuperjet.
\eitem

\bibliographystyle{plain}
\bibliography{biblio}

\end{document}